\documentstyle[latexsym,amssymb,aps,floats,eqsecnum,
preprint,amsfonts]{revtex}
\def\laq{\raise 0.4ex\hbox{$<$}\kern -0.8em\lower 0.62 ex\hbox{$\sim$}}
\def\gaq{\raise 0.4ex\hbox{$>$}\kern -0.7em\lower 0.62 ex\hbox{$\sim$}}
\input epsf
\begin{document}
\draft
\bibliographystyle{unsrt}


\title{Warped compactification on Abelian vortex in six dimensions}

\author{M. Giovannini\footnote{Electronic address: 
Massimo.Giovannini@ipt.unil.ch}, H. Meyer
\footnote{Electronic address:
 Harvey.Meyer@etu.unil.ch} and M. Shaposhnikov 
\footnote{Electronic address:  
Mikhail.Shaposhnikov@ipt.unil.ch }}

\address{{\it Institute of Theoretical Physics, 
University of Lausanne}}
\address{{\it BSP-1015 Dorigny, Lausanne, Switzerland}}

\maketitle

\begin{abstract} 

We consider the possibility of localizing gravity on a Nielsen-Olesen
vortex in the context of the  Abelian Higgs model. The vortex lives 
in a  six-dimensional space-time  with negative bulk cosmological
constant. In this model we find a region of the parameter space
leading, simultaneously, to warped compactification and to regular 
space-time geometry.  A thin defect limit is studied. Regular
solutions describing warped  compactifications in the case of higher
winding number are also  presented.

\end{abstract}

\vspace{3cm}

\noindent UNIL-IPT-01-4\\
hep-th/0104118\\
April 2001

\noindent
\newpage

\renewcommand{\theequation}{1.\arabic{equation}}
\setcounter{equation}{0}
\section{Introduction} 

Topological  defects appearing  in different  field-theoretical
models and residing  in higher-dimensional space-time can be 
considered as a prototype  of  four-dimensional  world  provided the 
known  particles  and gravity  are localized  on them 
\cite{m1}--\cite{rs}.  For  example, a  domain wall solution of a
simple $\lambda \phi^4$ theory with spontaneous breaking of $\phi
\rightarrow -\phi$ symmetry in five-dimensional space-time leads
naturally to  four-dimensional chiral fermions residing on  a domain
wall \cite{m1} and  to four-dimensional gravity \cite{rs},   if some
fine-tuning of  a bulk cosmological constant  to the domain  wall
tension is made. A number of  ``thick" wall solutions  in scalar
field theory  coupled to gravity  has  been  found   \cite{tw},  and 
the  possibility  of  gravity localization on these walls has been
studied.

If a domain wall is replaced by a more complicated
topological defect, such as  string \cite{def}-\cite{gs},  monopole
\cite{grs,def2,Dvali:2000ty} or  instanton 
\cite{Randjbar-Daemi:1983qa,Randjbar-Daemi:2000cr}  (living, respectively, 
  in  6, 7 or 8 dimensions), 
 the structure of chiral  fermionic zero modes, needed for
construction of a  realistic phenomenology  in four dimensions,  gets
richer. Hence, if the  possibility of constructing
standard model interactions  along these lines from higher 
dimensions is taken seriously, other fields (scalar,  gauge and
gravity) must be localized on topological defects as well.

In \cite{gs}  was shown that a thin  string in 6 dimensions could lead
to localization  of  gravity if  certain  relations  between the 
tension components are satisfied  (for even higher-dimensional
generalisations see  \cite{grs,Randjbar-Daemi:2000ft}).  The  aim 
of   the  present  paper  is  to  provide  a field-theoretical
realisation  of this idea  and show that an Abelian Higgs
model\cite{no}, in which  a string defect  arises naturally, can 
lead, at once,  to solutions which localize gravity and which have 
a geometry free of singularities. In the absence of branes a model
with an Abelian magnetic field in six dimensions was considered in
\cite{Gibbons:1987wg}.

The plan of our paper is the following. In Section II  the
Abelian-Higgs model coupled to six-dimensional Einstein-Hilbert
gravity is formulated. In Section III  the properties of 
six-dimensional cylindrically symmetric  warped geometries will be
discussed in  the  absence of defects.   This is  a necessary step
for  a classification of the solutions  in the presence of
vortex-type defects.  In Section IV the asymptotics of the solutions 
will be studied both for the metric and for the scalar and gauge
fields. Necessary conditions in order to get string solutions will
be  formulated. Section  VI contains  numerical  examples of  the 
solutions localizing  gravity whose parameter space  is analyzed  in
detail. The thin string limit is considered. In Section VII we 
discuss solutions leading  to gravity localization in the case  of
higher winding numbers and consider regular string solutions which do
not localize gravity. Section  VIII  contains our concluding remarks.
In the Appendix useful results concerning the interplay  between the
thin string limit and the Bogomolnyi limit have been collected.

\renewcommand{\theequation}{2.\arabic{equation}}
\setcounter{equation}{0}
\section{Abelian-Higgs model in six dimensions}
 \subsection{Basic equations}
The total action of gravitating Abelian Higgs model in six dimensions
can be written as 
\begin{equation}
S= S_{\rm brane} + S_{\rm grav}~,
\label{total}
\end{equation}
where $S_{\rm brane}$ is the gauge-Higgs action and $S_{\rm grav}$
is the six-dimensional generalization of Einstein-Hilbert gravity. 
More specifically,
\begin{equation}
S_{\rm brane}=\int
d^6x\sqrt{-G}{\cal L}_{\rm brane},~~
{\cal L}_{\rm brane}=
\frac{1}{2}({\cal D}_{A}\phi)^*{\cal D}^{A}\phi-\frac{1}{4}
F_{AB}F^{AB}
-\frac{\lambda}{4}\left(\phi^*\phi-v^2\right)^2~,
\label{a1}
\end{equation}
where ${\cal D}_{A}=\nabla_{A}-ieA_{A}$ is the gauge covariant derivative, 
while $\nabla_{A}$ is the generally covariant derivative 
\footnote{The conventions 
of the present paper are the following : the signature 
of the metric is mostly minus, Latin (uppercase) indices run over 
the $(4+2)$-dimensional space whereas Greek indices run 
over the four-dimensional space.}.
In Eq. (\ref{a1}), $v$ is the vacuum expectation value
 of the Higgs field
determining the masses of the Higgs and of the gauge boson 
\begin{equation}
m_{H} = \sqrt{2 \lambda}\, v,\,\,\,m_{V} = e\, v.
\end{equation} 
As usual the bulk action is 
\begin{equation}
S_{\rm grav}=-\int d^6x \sqrt{-G}
\left[\frac{R}{2\chi}+\Lambda\right],
\label{a2}
\end{equation}
where $\Lambda$ is a bulk cosmological constant, $\chi = 8\pi
G_6=8\pi/M_6^4$  and $M_{6}$ denotes the six-dimensional Planck
mass.

From Eqs. (\ref{total})--(\ref{a2}) the coupled system  of classical
equations of motion can be derived
\begin{eqnarray}
&&G^{A B}\nabla_{A}\nabla_{B}\phi
-e^2A_{A}A^{A}\phi-ieA_{A}\partial^{A}\phi-ie\nabla_{A}(A^{A}\phi)
+\lambda (\phi^*\phi-v^2)\phi=0~,
\label{ph}\\
&&\nabla_{A} F^{AB}=-e^2A^{B}\phi^*\phi+\frac{ie}{2}
\left(\phi\partial^{B}\phi^*-\phi^*\partial^{B}\phi\right)~,
\label{A}\\
&& R_{AB}-\frac{1}{2}G_{AB}R = \chi\left(T_{AB}+\Lambda G_{AB}\right)~,
\label{R}
\end{eqnarray}
where
\begin{equation}
T_{AB}=-{\cal L}_{\rm brane} 
G_{AB}\,+\, \frac{1}{2}\left[({\cal D}_A\phi)^*{\cal D}_{B}\phi 
+({\cal D}_B\phi)^*{\cal D}_A\phi\right]
-\,F_{AC}{F_B}^{C}.
\end{equation}

We are interested in static string-like solutions of Eqs. 
(\ref{ph})--(\ref{R})
that depend only on the extra coordinates and  thus do not break general
covariance along the four physical dimensions. For these solutions a
six-dimensional metric is \cite{m2,gs}
\begin{eqnarray}
ds^2=G_{AB} dx^{A} dx^{B} =
M^2(\rho)g_{\mu\nu}dx^\mu dx^\nu-d\rho^2-L(\rho)^2d\theta^2,
\label{metric}
\end{eqnarray}
where $\rho$ and $\theta$ are, respectively,  the bulk radius and the
bulk angle, $g_{\mu\nu}$ is the four-dimensional  metric and
$M(\rho)$, $L(\rho)$ are the warp factors. The Nielsen-Olesen ansatz
for the gauge-Higgs system reads:
\begin{eqnarray} 
&& \phi(\rho,\theta) =vf(\rho)e^{i\,n\,\theta},
\nonumber\\ 
&&A_{\theta}(\rho,\theta)  =\frac{1}{e}[\,n\,-\,P(\rho)] ~,
\label{NO}
\end{eqnarray}
where $n$ is the winding number.

Inserting Eq. (\ref{NO}) into Eqs. (\ref{ph})--(\ref{R})
and using the specific form of the  metric given in Eq. (\ref{metric}),
 the equations of motion can be written as
\begin{eqnarray} 
&& f'' + ( 4 m + \ell ) f'+(1 - f^2) f -
\frac{P^2}{{\cal L}^2}f=0,
\label{f1}\\ 
&& P'' + ( 4 m - \ell) P'  -\alpha f^2 P=0,
\label{p1}\\ 
&& \ell' + 3 m' + \ell^2 + 6 m^2 + 3 \ell m  = - \mu - \nu \tau_0 
+ \frac{\mu_{\rm ph}}{M^2},
\label{m1}\\
&& 4 m' + 10 m^2 = - \mu - \nu \tau_{\theta}
 + \frac{2~\mu_{\rm ph}}{M^2},
\label{m2}\\
&& 2 m \ell + 3m^2 = -\frac{\mu}{2} - \frac{\nu}{2} \tau_{\rho}
+\frac{\mu_{\rm ph}}{M^2}. 
\label{l1}
\end{eqnarray}
The four-dimensional metric $g_{\mu\nu}(\vec{x},t)$ obeys the equation:
\begin{equation} 
R^{(4)}_{\mu\nu}-\frac{1}{2}R^{(4)}g_{\mu\nu}=
\frac{8\pi}{M_P^2}\Lambda_{\rm ph} g_{\mu\nu}~,
\label{meti}
\end{equation}
where the physical cosmological term, $\Lambda_{\rm ph}$, is
an arbitrary  integration constant and $M_P$ is the four-dimensional
Planck mass. If $\Lambda_{\rm ph}=0$ we can
choose $g_{\mu\nu}=\eta_{\mu\nu}$ where  $\eta_{\mu\nu}$ is the
Minkowski metric. This will be the  case we will be mostly interested
in, even though we will often keep the physical cosmological
constant, for sake of completeness.

In Eqs. (\ref{f1})--(\ref{l1}) 
the prime denotes the derivation with respect to 
the rescaled variable  
\begin{equation}
x = \,{m_{H}\,\rho \over \sqrt{2}}~. 
\end{equation}
The function $L(\rho)$ appearing in the line element of Eq. 
(\ref{metric}) has also been rescaled, namely
\begin{equation}
{\cal L}(x) = \frac{m_{H}}{\sqrt{2}} L(\rho).
\end{equation}
Furthermore, the functions $m(x)$  and $\ell(x)$
are simply
\begin{equation}
m(x) = \frac{d \ln{M(x)}}{d x},\,\,\,
\ell(x) = \frac{d \ln{{\cal L}(x)}}{d x}.
\end{equation}
In Eqs. (\ref{f1})--(\ref{l1}) dimensionless parameters have been 
defined as:
\begin{equation} \alpha= 2 \frac{m^2_{V}}{m^2_{H}},~~~
\mu =\frac{2 \chi \Lambda}{ m^2_{H}},~~~ 
\nu= \chi \frac{m^2_{H}}{2\lambda},~~~ 
\mu_{\rm ph}=\frac{16\pi \Lambda_{\rm ph}}{M_P^2 m^2_{H}}. 
\label{constant}
\end{equation}

Within the mentioned conventions 
the components of the energy-momentum tensor
$T_{A}^{B}$ turn out to be
\begin{eqnarray} 
\tau_0(x)  &\equiv& T_{0}^{0}= T_{i}^{i}
= \frac{ {f'}^2}{2} + \frac{1}{4} ( f^2 -1 )^2 
+ \frac{{P'}^2}{2 \alpha {\cal L}^2 } + \frac{f^2 P^2 }{2
 {\cal L}^2},
\label{t0}\\ 
\tau_\rho(x)&\equiv& T_{\rho}^{\rho} =-\frac{{f'}^2}{2} 
+ \frac{1}{4} ( f^2 -1 )^2 - \frac{{P'}^2}{2 \alpha {\cal L}^2 } 
+ \frac{f^2 P^2 }{2 {\cal L}^2},
\label{trho}\\ 
\tau_\theta(x)&\equiv& T_{\theta}^{\theta}=
 \frac{{f'}^2}{2} + \frac{1}{4}  ( f^2 -1 )^2 
- \frac{{P'}^2}{2 \alpha {\cal L}^2 } - \frac{f^2 P^2  }{2 {\cal L}^2}.
\label{tth} 
\end{eqnarray}
Eqs. (\ref{f1})--(\ref{p1}) correspond, respectively, to  the
equations of motion for the scalar and gauge fields whereas Eqs.
(\ref{m1})--(\ref{l1}) correspond to the  various components of
Einstein's equations. Eqs. (\ref{m1})--(\ref{m2}) are truly dynamical
since they involve $m'(x)$ and $\ell'(x)$.  Eq. (\ref{l1}) connects
$m(x)$ and  $\ell(x)$ to the first derivatives of the gauge and
scalar degrees of freedom and it is, therefore, a constraint.
In the following part of the paper $\mu_{\rm ph}=0$ will be assumed.

\subsection{Boundary conditions}
Regular solutions of the equations of motion can be investigated once 
 the boundary
conditions are specified. In order to describe a string-like
defect in six  dimensions we have to demand that the scalar field 
reaches, for large $\rho$, its vacuum expectation  value, namely
$|\phi(\rho)| \rightarrow v$  for $\rho \rightarrow \infty$. In the
same limit, the magnetic field should go to zero. Moreover, close to
the core of the string  both fields should be regular. These
requirements  can be translated in terms of our  rescaled variables as
\begin{eqnarray} 
f(0)=0,&\qquad& \lim_{x\rightarrow \infty} f(x)=1,
\nonumber\\ 
P(0)=n,&\qquad& \lim_{x\rightarrow \infty} P(x)=0. 
\label{boundary}
\end{eqnarray}
The requirement of regular geometry in the core of the string reads:
\begin{equation}
M'(0) = 0~,~~{\cal L}(0) = 0~,~~ {\cal L}'(0) = 1~,
\label{boundary2}
\end{equation}
and one can arbitrarily fix $M(0) = 1$. 

These conditions listed in Eqs. (\ref{boundary})--(\ref{boundary2}) 
completely specify a solution. Clearly, Eqs. 
(\ref{boundary})--(\ref{boundary2})  do not tell
anything about the asymptotics of the metric at large distances from
the string and thus a possibility of gravity localization on a
defect remains open. The requirement of gravity localization is
equivalent to the finiteness of the four dimensional Planck mass,
\begin{equation}
M_{P}^2 = \frac{4 \pi M^4_{6}}{m_{H}^2} \int dx M^2(x) {\cal L}(x) 
< \infty~,
\label{4dpl}
\end{equation}
which does not hold in general and requires a fine-tuning of parameters
of the model, as it will be discussed.

\renewcommand{\theequation}{3.\arabic{equation}}
\setcounter{equation}{0}
\section{Bulk solutions} 
Outside the core of the string all source terms in Eqs. 
(\ref{t0})--(\ref{tth}) vanish and a general solution to equations
\begin{eqnarray}
&&\frac{\partial m}{\partial x} + \frac{5}{2} m^2 
+ \frac{\mu}{4}=0,
\nonumber\\
&&  \ell=-\frac{\mu}{4m}-\frac{3m}{2},
\label{vaceq}
\end{eqnarray}
can be easily found \cite{m2}. We will consider the case $\Lambda \leq
0$ only, since the case of a 
 positive cosmological constant in the bulk was studied in
\cite{m2}. Defining 
\begin{equation}
c = \sqrt{- \frac{\mu}{10}} > 0,
\end{equation}
the solutions of Eqs. (\ref{vaceq}) can be written as 
\begin{equation}
m(x)=- c~\frac{1-\epsilon\,e^{ 5 c x }}
{1+\epsilon e^{ 5 c x}}\,,
\label{mm}
\end{equation} 
where $\epsilon$ is an integration constant. Once $m(x)$ is known,
$\ell(x)$ can be determined  from the second of Eqs. (\ref{vaceq}).

In order to understand whether gravity can be localized on any of
those solutions we compute $M$ and ${\cal L}$:
\begin{eqnarray}
&& M(x) = M_0 e^{-cx} |1+\epsilon e^{5cx}|^{2\over 5}\, ,
\nonumber\\
&& {\cal L}(x) = {\cal L}_0 e^{-cx}
\frac{|\epsilon e^{5cx} - 1|}{|1+\epsilon e^{5cx}|^{3\over 5}}\, .
\label{epsarb}
\end{eqnarray}

If $\epsilon > 0$ or $\epsilon \leq -1$ we have that for $x \rightarrow
\infty$ $M\sim {\cal L} \sim e^{cx}$. For this solution the integral
in (\ref{4dpl}) diverges and thus gravity cannot be localized. 

If $\epsilon=0$ the solution is simply 
\begin{equation}
m(x) = \ell(x) = -c,
\label{eps=0}
\end{equation}
and the warp factors
will be exponentially  decreasing as a function of the bulk radius:
\begin{eqnarray}
&& M(x) = M_0 e^{-c x} ,\,\,\,
\nonumber\\
&& {\cal L}(x) = {\cal L}_0 e^{-c x}.
\label{LM}
\end{eqnarray}
The solution of Eqs. (\ref{eps=0})--(\ref{LM}) 
 leads to gravity localization and to a
smooth AdS geometry far from the string core. We also note that a
pure positive exponential solution can be derived for $\epsilon
\rightarrow \pm \infty$ taking $M_0 \sim {\rm c}_1/|\epsilon|^{2\over5}$ and
$ {\cal L}_0 \sim {\rm c}_2/|\epsilon|^{2\over5}$ where ${\rm c}_1$ and 
${\rm c}_2$ are two arbitrary constants.

If $-1<\epsilon < 0$, $M(x_0)=0$ where $x_0=
\frac{1}{5c}\log{1/|\epsilon}|$. The geometry is singular at
$x=x_0$ (see below) and  $x<x_0$ should be required. In spite of the fact
that ${\cal L}$ diverges at $x=x_0$, the integral (\ref{4dpl})
defining the four dimensional Planck mass is finite. So, these geometries
can be potentially used for a gravity localization, provided the
singularity at $x_0$ is resolved by some way, e.g. by string theory \cite{max}.
These types of singular solutions were discussed in \cite{m2} for the case 
of positive cosmological constant. 

Finally, if the  bulk cosmological constant is zero,  the solutions have a
power-law behaviour,  namely
\begin{equation}
M(x) \sim x^{\gamma},\,\,\,{\cal L}(x) \sim x^{\delta},
\end{equation}
with 
\begin{equation}
d \gamma + \delta=1, \,\,\,\, d \gamma^2 + \delta^2 =1 \, ,
\end{equation}
where $d=4$ is the number of dimensions of the metric $g_{\mu\nu}$. 
These solutions belong to the  Kasner class. The
Kasner conditions leave open only two possibility: either $\delta =
1$ and $\gamma = 0$ or $\gamma= 2/5$ and $\delta = -3/5$, [the
same fractions as in Eqs. (\ref{epsarb})]. None of them
could lead to localization of gravity.

In order to study the singularity structure 
 of the bulk  solutions without specifying $\epsilon$, the different
curvature invariants may be computed. 
In the case of warped metrics the Riemann
invariant has been calculated  in \cite{ran}. The explicit form of all
the  curvature invariants for the metric of Eq. (\ref{metric}) is
given  in \cite{max} and they are:
\begin{eqnarray}
R&=&20m^2+8m'+2\ell'+2\ell^2+8\ell m,
\label{scalarc}\\
R_{AB}R^{AB}&=&80m^4+20m'^2+2\ell'^2+2\ell^4+64m'm^2+
4\ell^2\ell'+28m^2\ell^2
\nonumber\\
&+&32m^3\ell +8mm'\ell+8m'\ell'+8m'\ell^2+8m^2\ell'+8m\ell\ell'+
8m\ell^3,
\label{ricci}\\
R_{ABCD}R^{ABCD}&=&4\ell^4-24m^4+8\ell^2\ell'
+4\ell'^2+32m^2m'+16m'^2,
\label{riemann}\\
C_{ABCD}C^{ABCD}&=&\frac{12}{5}
\left[(m'-\ell')+\ell(m-\ell)\right]^2.
\label{weyl}
\end{eqnarray}

Let us first check the regularity properties  of the bulk solutions
(\ref{mm}). The Ricci and scalar curvature invariants are simply 
constant for any $\epsilon$:
\begin{equation}
R= 30 c^2,\,\,\, R_{AB}R^{AB} = 150 c^4.
\end{equation}
The Riemann and Weyl invariants are
\begin{eqnarray}
&& R_{ABCD}R^{ABCD}=
\frac{-20\,c^4\,\left( e^{- 20\,c\,x} - 12\,e^{- 15\,c\,x}\,
\epsilon - 
      138\,e^{- 10\,c\,x}\,\epsilon^2 - 12\,e^{- 5\,c\,x}\,
\epsilon^3 + \epsilon^4 \right)
      }{{\left( e^{- 5\,c\,x} + \epsilon \right) }^4},
\nonumber\\
&& C_{ABCD}C^{ABCD}=
\frac{3840\,c^4\,e^{-10\,c\,x}\,\epsilon^2}{{\left( e^{- 5\,c\,
x} + \epsilon \right) }^4}.
\end{eqnarray}
From these last two expressions we can see that if $\epsilon > 0$ or
$\epsilon < -1$ the Weyl and Riemann invariants  do not have any pole
and thus we have a regular geometry. For $\epsilon = 0$ all curvature
 invariants
are constant and we have a six-dimensional 
AdS space.  On the other hand if $-1 <
\epsilon < 0$  both Weyl and Riemann invariants diverge at a finite value  of
$x$  leading to a singular geometry. The case  $\epsilon
=-1$ is somewhat specific since 
a singularity of Kasner
type is developed in the origin. 

The parameters $\epsilon$, $M_0$, ${\cal L}_0$ are not fixed for a
bulk space solution. If a string-like defects
is placed at the origin $x=0$ these constants are
no longer arbitrary and are functions of the three parameters of
the model, namely, $\epsilon = \epsilon(\alpha,\mu,\nu)$, and so on. 
An everywhere regular geometry  is simultaneously  achieved together 
with gravity localization if the parameters 
lie on the surface $\epsilon(\alpha,\mu,\nu)=0$.

\renewcommand{\theequation}{4.\arabic{equation}}
\setcounter{equation}{0}
\section{Asymptotics of the solutions}
 
\subsection{The asymptotics of the solutions at the origin}

The form of the solutions in the vicinity of the core of the vortex
can be  studied by  expressing the metric functions together with
the  scalar and gauge fields  as a power series in $x$, the
dimensionless bulk radius.  The power series will then be inserted
into Eqs.  (\ref{f1})--(\ref{l1}). Requiring that the series obeys,
for $x\rightarrow 0$,  the boundary conditions of Eqs.
(\ref{boundary})  the form of the solutions can be determined as a
function of the  parameters of the model. 

Consider, in particular, the case of $n=1$ \footnote{This  analysis can be
generalized to the case of higher winding number  (namely $n \geq
2$) as it will be seen in Section VI.}. 
In this case  asymptotic solutions close to the core can be  written
as :
\begin{eqnarray}
f(x)&\simeq&Ax+\left(\frac{2\mu}{3}+\frac{\nu}{6}+\frac{4B^2 
\nu}{3\alpha}+\frac{2A^2 \nu}{3}-1+2B\right)\frac{A}{8}x^3,\\
P(x)&\simeq&1+Bx^2,\\
M(x)&\simeq&1+\left(-\frac{\mu}{8}-\frac{\nu}{32}+
\frac{\nu B^2}{4\alpha}\right)x^2,\\
{\cal L}(x)&\simeq&x+\left[\frac{\mu}{12}+\nu(\frac{1}{48}
-\frac{5B^2}{6\alpha}-\frac{A^2}{6})\right]x^3.
\label{x=0}
\end{eqnarray}
In Eq. (\ref{x=0}) $A$ and $B$ are two arbitrary constants which cannot 
be determined by the local analysis of the equations of motion. These
constants are to be found by boundary conditions for $f(x)$ and
$P(x)$ at infinity. 

For completeness, we give also the  limit 
of the  components of the energy-momentum tensor in the vicinity 
of the origin:
\begin{eqnarray}
&&\tau_{0} \simeq \biggl[ \frac{1}{4} +  A^2 
+ \frac{2\,B^2}{\alpha}\biggr],
\nonumber\\
&& \tau_{\rho} \simeq \biggl[ \frac{1}{4} 
- \frac{ 2 B^2}{\alpha} \biggr],
\nonumber\\
&& \tau_{\theta} \simeq \biggl[ \frac{1}{4} 
- \frac{ 2 B^2}{\alpha} \biggr].
\label{tx=0}
\end{eqnarray}

\subsection{The asymptotics of the solutions at infinity}
Consider now asymptotics of solutions at large $x$, assuming that the
geometry is regular at infinity. This situation can be realized 
both in the case  $\epsilon =0$ [i.e. Eq. (\ref{eps=0})] and  in the case 
 Eq. (\ref{epsarb}) either with $\epsilon > 0$ or with 
$\epsilon < -1$. We can write 
\begin{eqnarray}
&& P(x) = \overline{P} + \delta P(x),
\nonumber\\
&& f(x) = \overline{f} - \delta f(x),
\end{eqnarray}
where, according to Eqs. (\ref{boundary}) for $x\gg 1$, $\overline{P}
\sim 0$ and $\overline{f} \sim 1$.  

First, we will consider the case 
$\epsilon =0$. From Eq. (\ref{p1}) it follows
that 
\begin{equation}
\delta P'' - 3 c \delta P' - \alpha \delta P=0,
\label{eqp}
\end{equation}
which  implies that, for $x\gg 1$
\begin{equation}
\delta P(x) \sim e^{ \sigma_{1} x},\,\,\sigma_1 = 
\frac{3 c}{2} \bigl[ 1 \pm \sqrt{1+  \frac{4 \alpha}{9
c^2}}\,\,\bigr]~.
\label{dP}
\end{equation}
If  $4\alpha \gg 9 c^2$ (the limit of small bulk cosmological
constant) the solution  is compatible with the gauge field decreasing
asymptotically as  $\delta P \sim e^{-\sqrt{\alpha}x}$ which is the
well  known flat-space result of the Abelian-Higgs model.

The linearization of the scalar field equation, using a similar 
procedure, yields
\begin{equation}
\delta f'' - 5 c \delta f' - 2 \delta f =0,
\label{eqf}
\end{equation}
leading to the solution
\begin{equation}
\delta f(x) \sim e^{\sigma_2 x},\,\,\,\, \sigma_2 = 
\frac{5 c}{2} \bigl[ 1 \pm \sqrt{ 1 + \frac{8}{25 c^2}}\,\,\bigr]~.
\label{df}
\end{equation}
If  $ 25 c^2 \ll 8$ the perturbed  solution goes as  $\delta f \sim
e^{- \sqrt{2} x}$ which, again, is the  flat-space result. 

In Eqs. (\ref{dP})--(\ref{df}) a minus sign should be chosen
to obtain an exponentially decreasing behaviour. Moreover, linear
approximation is only valid if $\sigma_2 > 2\sigma_1 + 2c$, 
in the opposite case $\delta f \sim \exp(2\sigma_1 + 2c)$.

For $\epsilon \neq 0$ the sign in front of $c$ in equations
(\ref{eqp},\ref{eqf}) must be changed, and we get
\begin{eqnarray}
&& \delta P(x) \sim e^{ \sigma_{1} x},\,\,\sigma_1 = 
-\frac{3 c}{2} \bigl[ 1 +\sqrt{1+  \frac{4 \alpha}{9 c^2}}\,\,\bigr]~,
\nonumber\\
&&\delta f(x) \sim e^{\sigma_2 x},\,\,\,\, \sigma_2 = 
-\frac{5 c}{2} \bigl[ 1 + \sqrt{ 1 + \frac{8}{25 c^2}}\,\,\bigr]~.
\label{ris} 
\end{eqnarray}
The linear
approximation is valid if $\sigma_2 > 2\sigma_1 - 2c$, otherwise 
 $\delta f \sim \exp(2\sigma_1 - 2c)$.

\subsection{String tensions} 

In the description  of string-like defects in four-dimensions  it is
useful to study the relations obeyed by the string tensions. 
Indeed, some particular linear combinations of the string tensions 
may be related to the Tolman mass per unit length and to the  angular
deficit of the geometry. These quantities can be used  in order to
classify the physical  properties of the solutions and in order to
check if they  are effectively  asymptotically conical \cite{abh}. 

This analysis is also relevant in our case even though the physical 
interpretation  changes and the specific algebraic relations are
different from the  well  known four-dimensional cases. The
components of the string tension are defined as
\begin{equation}
\mu_i = \int_0^\infty dx M^4(x){\cal L}(x)\tau_i(x).
\label{tension}
\end{equation} 
The integrals appearing at the right hand side always converge for 
solutions with regular geometry, as follows from the results of the
previous subsection. 

We will show now that for $M(x) \rightarrow 0,~~{\cal L}(x)
\rightarrow 0$  at $x \rightarrow \infty$ (i.e. in the case we are
mostly interested in) the constant $B$ in Eq. (\ref{x=0}) can be
analytically determined.

Consider a specific linear combination of Eqs.
 (\ref{m1})--(\ref{l1}), namely
\begin{eqnarray}
&&2m' + 4 m^2 + m\ell =-\frac{\mu}{2}-\frac{\nu}{4}\left[  
\tau_\rho
+ \tau_\theta \right],
\label{c1}\\
&& \ell' + \ell^2 + 4\ell m 
=-\frac{\mu}{2}-\nu\left[  \tau_0+\frac{1}{4} \tau_\rho
-\frac{3}{4} \tau_\theta \right].
\label{c2}
\end{eqnarray}
Integrating Eqs. (\ref{c1})--(\ref{c2}) from zero to $x_c \to \infty$,
we get
\begin{eqnarray}
&& M^3(x_c)M'(x_c){\cal L}(x_c)=-\frac{\mu}{2}\int_0^{x_c} 
M^4 {\cal L}dx
-\frac{\nu}{4}\left(\mu_\rho+\mu_{\theta}\right),
\label{rela}\\ 
&&M(x_c)^4{\cal L}'(x_c)=1-\frac{\mu}{2}\int_0^{x_c} M^4 {\cal L}dx-\nu
\left(\mu_0-\frac{3}{4}\mu_{\theta}
+\frac{\mu_\rho}{4}\right).
\label{relb} 
\end{eqnarray} 
In the limit $x_c \to \infty$, Eq. (\ref{rela}) is the 
six-dimensional analog  of the relation determining the Tolman mass
whereas Eq. (\ref{relb})  is the generalization of the relation
giving the angular deficit. Therefore the following relation must
hold: 
\begin{equation} 
\mu_0 -\mu_{\theta}=\frac{1}{\nu}\,.
\label{mus} 
\end{equation} 
This is the fine-tuning relation which has been already 
found in \cite{gs,grs}. 

Let us note that if $\epsilon \neq 0$ one can
use the asymptotics of the metric found in
Eq.(\ref{epsarb}) for computation of the left-hand sides of
(\ref{rela},\ref{relb}). This gives a more general relation,
\begin{equation} 
\mu_0 -\mu_{\theta}=\frac{1}{\nu}(1 - 10\,c\,\epsilon\, M_0^4\,{\cal
L}_0)~.
\label{musgen} 
\end{equation} 
The extra term vanishes also for $\epsilon \to \pm \infty$ given the
scaling behaviour of $M_0^4 \sim |\epsilon|^{-\frac{8}{5}}$ 
and ${\cal L}_0\sim |\epsilon|^{-\frac{2}{5}}$. So, the condition (\ref{mus})
is only necessary but not sufficient in  order to have solutions leading to
gravity localization.

Being an integral relation  we can use (\ref{mus}) in order to get
informations  on the behaviour of the solution  in the origin.  In
order to do so, the left hand side of  Eq.
(\ref{mus})  can be directly expressed 
in terms of the scalar and vector fields 
\begin{equation}
\mu_0-\mu_\theta= \int M^4 {\cal L}
\left[\frac{P'^2}{\alpha{\cal L}^2}
+\frac{f^2P^2}{{\cal L}^2}\right] dx~.
\label{in}
\end{equation}
Using now Eq. (\ref{p1})  we arrive at
\begin{equation}
\left(\frac{M^4P'}{{\cal L}}\right)'=\alpha\frac{f^2PM^4}{\cal L},
\end{equation}
which can be inserted back into Eq. (\ref{in}). Integrating 
by parts the obtained relation the term containing 
${P'}^2$ cancels and we 
arrive at
\begin{equation}
\mu_0-\mu_\theta=\frac{1}{\alpha}\biggl(
\left.\frac{P\,P'\,M^4}{\cal L}\right|_{\infty}
-\left.\frac{P\,P'\,M^4}{\cal L}\right|_0\biggr)~.
\label{in2}
\end{equation}

For the solutions we are interested in the boundary term at infinity vanishes. 
Using now the asymptotic behaviour 
for small $x$ of Eqs. (\ref{x=0}) we obtain, from Eq. (\ref{in2}),
\begin{equation}
\mu_0-\mu_\theta=
-\frac{1}{\alpha}\left.\frac{P'}{\cal L}\right|_0.
\label{in3}
\end{equation}
In this case, from Eq. (\ref{mus}) and using Eq. (\ref{in2}) we 
have that
\begin{equation}
-\frac{\nu}{\alpha}\left.\frac{P'}{\cal L}\right|_0=1\,.
\label{mus2}
\end{equation}
According to Eq. (\ref{x=0}), for $x \rightarrow 0$, 
$ P\sim 1 + B x^2$. Using 
Eq. (\ref{mus2}) the expression for $B$ can be exactly computed
\begin{equation}
B= - \frac{\alpha}{2\nu}\,.
\label{B}
\end{equation}
This relation is very helpful for numerical integration of equations
of motion performed in the following section.

Finally,  we would  like to  notice that  the magnetic  field  can be
expressed  as   ${\cal  B}(\rho)=-P'(\rho)/(e\,L(\rho))$.   Using  now
Eq. (\ref{B}) we have that $8 \pi {\cal B}(0) =e\,M_6^4$.

In closing the present Section we want to stress that the results 
obtained from the analysis of the relations among the tensions 
are only necessary conditions in order to find $\epsilon=0$ solutions.
They are not sufficient because $f'(0) \sim A$ still remains
undetermined.

\renewcommand{\theequation}{5.\arabic{equation}}
\setcounter{equation}{0}
\section{Solutions with localization of gravity}
In this Section, a numerical search for solutions that
localize gravity will be performed. 
The general method of integration will be firstly outlined. 
Secondly,  some  interesting numerical solutions will be 
presented as an example.
Finally, the scan of the parameter space of the 
model leading to $\epsilon(\alpha,\mu,\nu) =0$ will be presented.

\subsection{Numerical integration}

In order to prepare the ground for the numerical integration 
the system of Eqs. (\ref{f1})--(\ref{l1}) can be expressed in a form 
of a  first order 
non-linear dynamical system.
By linearly combining Eqs. (\ref{f1})--(\ref{l1}) the following
set of equations can be obtained:
\begin{eqnarray}
&&f'=g,
\nonumber\\
&&P'=Q,
\nonumber\\
&&{\cal L}'=l,
\nonumber\\
&&g'=-\left(4m+\frac{l}{{\cal L}}\right)g-(1-f^2)f
+\frac{P^2}{{\cal L}^2}f,
\label{I}\\
&&Q'=-\left(4m-\frac{l}{{\cal L}}\right)Q+\alpha Pf^2,
\label{II}\\
&& m'=-\frac{5}{2}m^2-\frac{\mu}{4}-\frac{\nu}{4}\left(\frac{g^2}{2}
+\frac{1}{4}(1-f^2)^2-\frac{Q^2}{2\alpha {{\cal L}}^2}
-\frac{f^2P^2}{2{\cal L}^2}\right),
\label{III}\\
&& l'=-\frac{\mu}{4}{{\cal L}} +\frac{3}{2}m^2 {{\cal L}}
-3lm-\frac{\nu {\cal L}}{4}\left( \frac{g^2}{2}
+\frac{1}{4}(1-f^2)^2+\frac{7Q^2}{2\alpha {\cal L}^2}
+\frac{7f^2P^2}{2{\cal L}^2}\right),
\label{IV}\\
&&M'=m M.
\label{V}
\end{eqnarray}
In order to solve this system the shooting method \cite{num} 
has been employed. 
In the shooting method a boundary value problem is tackled as a series of 
 Cauchy problem. Using forward integration in $x$ with a specific set 
of initial conditions, the asymptotic values of the fields are found. 
If 
the asymptotic values are different from the boundaries discussed 
in Eq. (\ref{boundary}), the integration is performed again with a
different set of initial conditions until the correct asymptotic 
boundary is reached.

The technical problem is twofold. On one hand the shooting method 
requires that the initial conditions are 
given in a range of the parameter space sufficiently close 
to the one admitting $\epsilon(\alpha,\mu,\nu)=0$ 
solutions. On the other hand,  
for the scalar and gauge fields  the integration is
a boundary value problem whereas, for the warp factors,  
the integration is a Cauchy problem. 
The shooting method is implemented in such a way that
each   forward integration is performed by a Runge-Kutta routine. 
The differences between the obtained values at large 
$x$ and the boundary values of Eq. (\ref{boundary}) are 
 considered as functions
 of  the constants $A$ and $B$ appearing in the asymptotic 
solution in the vicinity of the core. 
The Newton method \cite{num} is  then used in order to find 
the values of $A$ and $B$  required in order to match the boundaries 
within the precision of the algorithm.

Therefore, in order to identify a correct set of initial conditions 
extensive use of backward integrations has been made. Backward integrations 
have been applied by giving 
final conditions on warped solutions at large $x$ and by 
integrating the equations from large to small $x$.  This method 
allowed to 
identify a set of initial conditions leading to the correct asymptotic 
boundaries with warped metrics. This idea has also been applied 
in a different context always involving warped 
compactification in string inspired models with quadratic 
curvature corrections \cite{max}.

\subsection{Some numerical examples}

\begin{figure}
\centerline{\epsfxsize = 11 cm  \epsffile{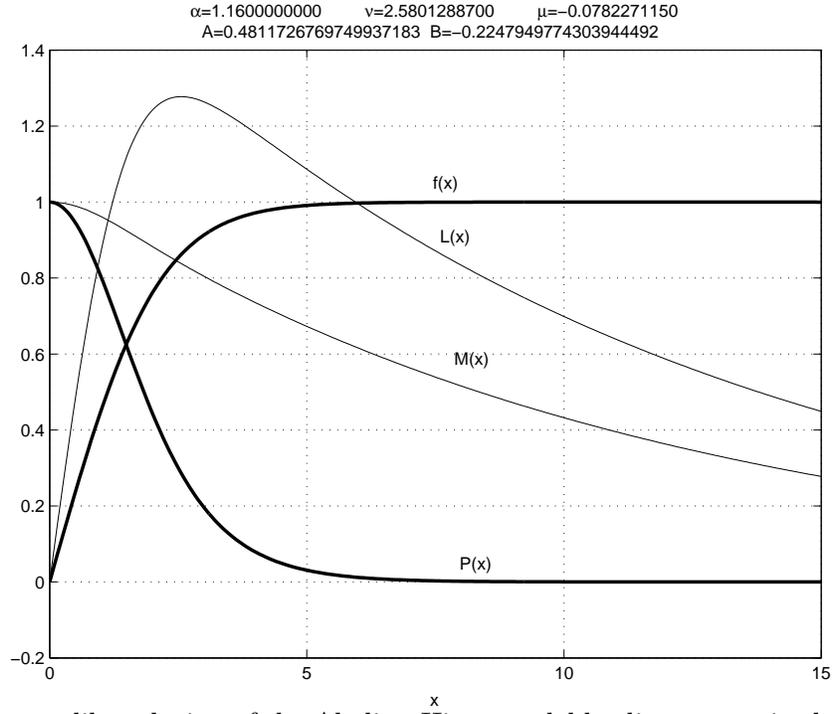}} 
\caption[a]{A vortex-like solution of the Abelian-Higgs model leading 
to gravity localization. With the full thick lines we report
the behaviour of the scalar and gauge fields whereas with 
the thin lines the behaviour of the warp factors is illustrated. The same 
convention will be used throughout the paper.}
\label{F4}
\end{figure}

We are now going to give some examples of the numerical 
integration, see Fig. \ref{F4}. The values of the 
scalar and gauge fields are the ones dictated by Eqs. (\ref{boundary}).
At the core of the vortex, 
the behaviour of the metric and of the other fields follows  
Eqs. (\ref{x=0}). For large $x$ the metric functions 
decay exponentially. The asymptotic behaviour 
of the metric functions for large $x$ 
belongs to the class of solutions discussed in Section II for 
$\epsilon =0$, namely $m \sim \ell \sim -\sqrt{-\mu/10}$. 

\begin{figure}
\centerline{\epsfxsize = 11 cm  \epsffile{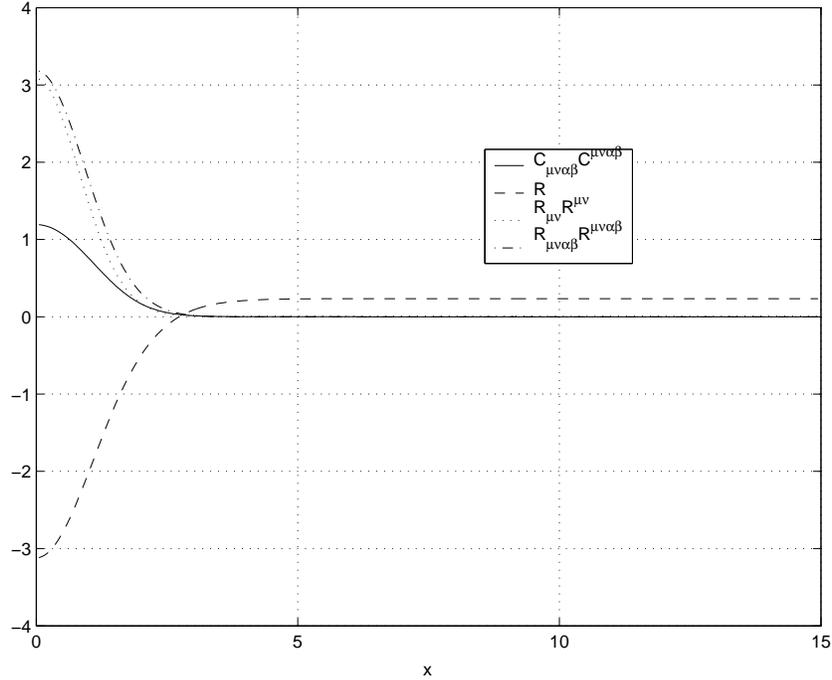}} 
\caption[a]{The curvature invariants for the solution 
shown in Fig. \ref{F4}.}
\label{F5}
\end{figure}

\begin{figure}
\centerline{\epsfxsize = 11 cm  \epsffile{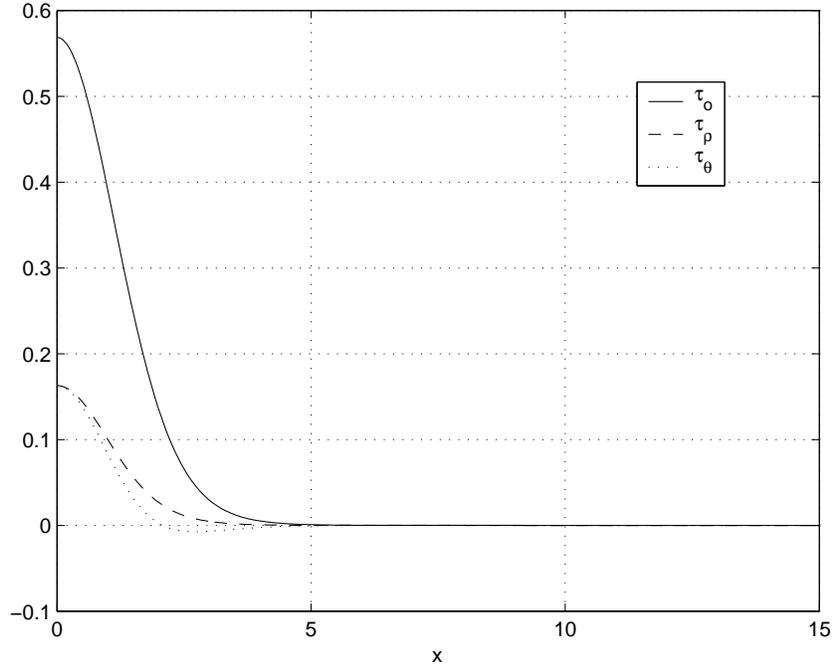}} 
\caption[a]{The components of the energy momentum 
tensor of the sources in the case of the numerical 
solution described in Fig. \ref{F4}.}
\label{F6}
\end{figure}

In the example of Fig. \ref{F4},
$\alpha = 1.16$. The relationship (\ref{B}) is satisfied.
The curvature invariants 
(scalar curvature, Riemann, Weyl and Ricci invariants)
defined in Eqs. (\ref{scalarc})--(\ref{weyl})  are plotted in Fig. \ref{F5}
for the solution illustrated in Fig. \ref{F4}.
The Riemann and Ricci 
invariants go asymptotically to a constant which is ${\cal O}(c^4)$, whereas 
the scalar curvature goes to a constant ${\cal O}(c^2)$ as 
dictated by the 
asymptotic 
solutions with $\epsilon =0$ and derived in Eq. (\ref{eps=0}). 
The Weyl invariant goes to zero since, asymptotically, 
the space-time is isotropic.
Finally it is interesting to check the behaviour of the 
various components 
of the energy momentum tensor, which are reported in Fig. \ref{F6}. 
From Fig. \ref{F6} we notice that $\tau_{\theta}(x)$ gets negative 
and it approaches zero from negative values. This 
is due to the fact that, in $\tau_{\theta}(x)$ 
the magnetic stresses dominate leading to locally 
negative pressure.

\begin{figure}
\centerline{\epsfxsize = 11 cm  \epsffile{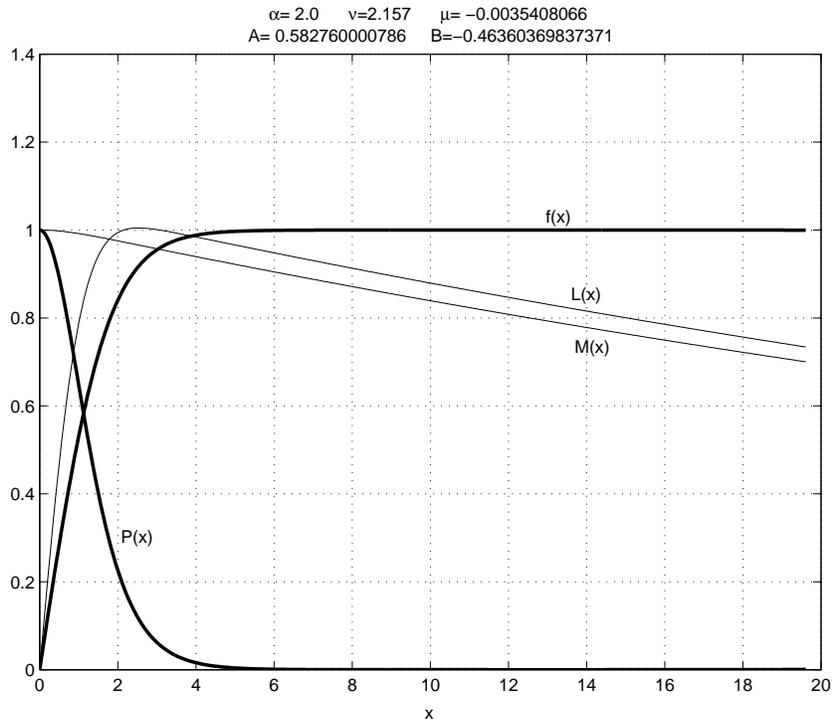}} 
\caption[a]{ A warped  solution is illustrated in the 
case $\alpha=2$ where $m_{H} = m_{V}$.}
\label{F7}
\end{figure}

In Fig. \ref{F7}, we show an example where the exponential damping of 
the metric is slower than in the case showed in Fig. \ref{F4}. 
The reason 
for this behaviour stems from the fact that 
the  value of the cosmological constant (i.e. the $\mu$ parameter)
chosen in Fig. \ref{F7} 
is roughly one order of magnitude smaller than in the case 
discussed in Fig. \ref{F4}. The case  $\alpha=2$ corresponds to 
$m_{V} \equiv m_{H}$, i.e. Bogomolnyi limit where the Abelian-Higgs 
model in four dimensional flat space obeys interesting properties.
This limit is discussed in some details in the Appendix.
To check for the accuracy of numerical integration 
procedure we consider the constraint equation (\ref{l1}) 
(which was not used in the numerical procedure) that connects 
$m(x)$  and $\ell(x)$ 
to the $\tau_{\rho}(x)$ component of the energy-momentum tensor:
\begin{equation}
2 m \ell + 3m^2 +\frac{\mu}{2} + \frac{\nu}{2} \tau_{\rho}
=0.
\label{c}
\end{equation} 
Inserting the numerical solutions obtained in the present 
Section into Eq. (\ref{c})  we should find that Eq. (\ref{c}) 
is numerically satisfied. The accuracy for which 
Eq. (\ref{c}) is satisfied indicates the accuracy 
of the integration.

\begin{figure}
\centerline{\epsfxsize = 11 cm  \epsffile{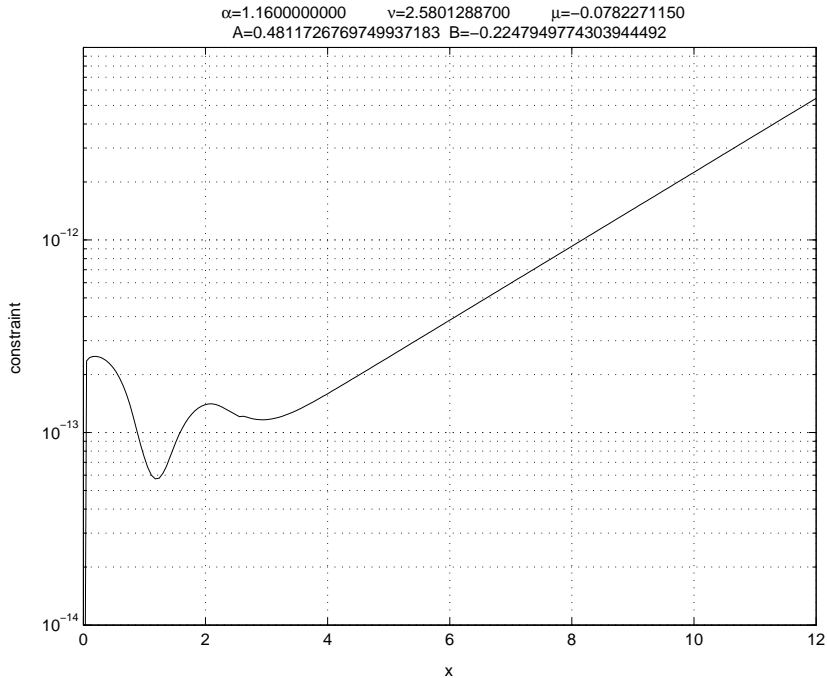}} 
\caption[a]{The constraint Eq. (\ref{c}) is 
illustrated for the case of the solution given in Fig. \ref{F4}.}
\label{F7b}
\end{figure}

In Fig. \ref{F7b} we plot the left hand side of Eq. (\ref{c}) 
for the solution shown in Fig. \ref{F4} of this Section.
As we can see Eq. (\ref{c}) is satisfied with a precision 
of $10^{-13}$ in the vicinity of the core and with a precision
of $10^{-11}$ for large $x$. 
For the other solutions similar behaviour can be obtained
for the constraint.

\subsection{ The parameter space of the solutions 
with gravity localization}

Up to this moment, examples of compactifications on a 
vortex were analyzed. A scan in 
the parameter space of the solution will now be performed.

In order to achieve this program 
 in a systematic way the procedure is, in short, the following.
Suppose that we have already a solution that satisfies all the 
requirements (for example, the one in Fig. \ref{F4} or \ref{F7}).
It can be found by a
procedure of backward integration discussed above. 
Then one of the three parameters among $\alpha$, $\nu$ and $\mu$ 
 is kept fixed, a second is changed by small steps 
and the third one is tuned to the first two by requiring that
\begin{eqnarray}
&&\ell(x)+ \sqrt{\frac{-\mu}{10}} =0,
\nonumber\\
&& m(x) + \sqrt{\frac{-\mu}{10}} =0,
\label{zero}
\end{eqnarray}
are satisfied for large $x$. The zeroes of Eq. (\ref{zero}) 
are found by
means of a simple bisection method \cite{num}. 

\begin{figure}
\centerline{\epsfxsize = 11 cm  \epsffile{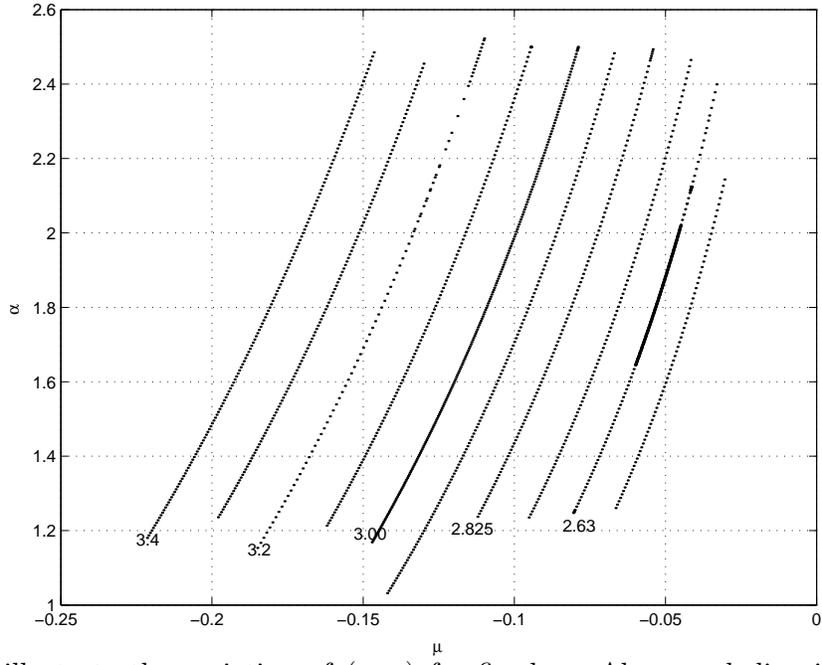}} 
\caption[a]{We illustrate the variation of $(\alpha,\mu)$ 
for fixed $\nu$. Along each line in this plane the numerical 
$\nu$ is constant and it is given by the numerical values 
reported at the bottom of the curves. }
\label{F9}
\end{figure}

Using this procedure a scan of the parameter space of the 
model has 
been performed. Let us start by discussing the scan in the 
$(\mu,\alpha)$
plane at constant $\nu$.  
The results of the study are shown in 
Fig. \ref{F9}.

\begin{figure}
\centerline{\epsfxsize = 11 cm  \epsffile{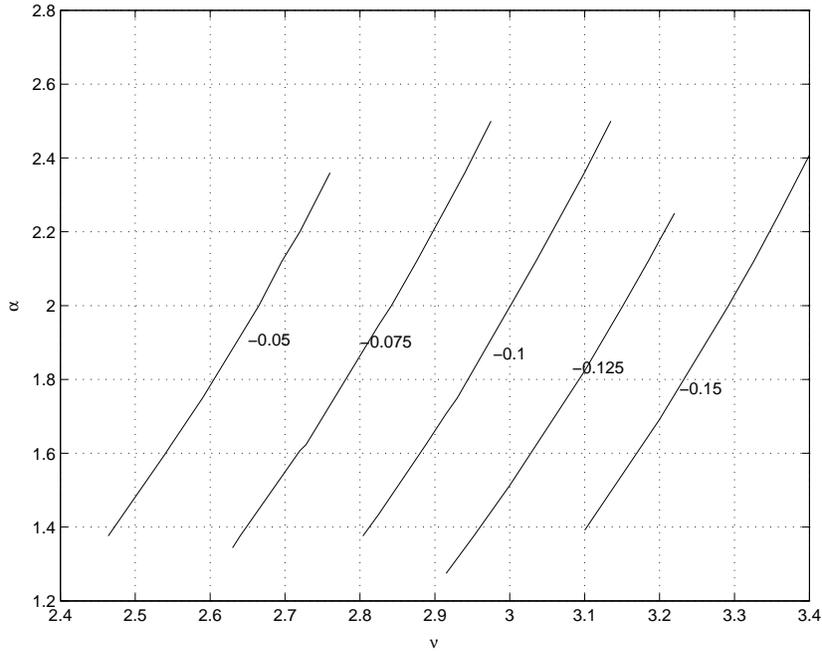}} 
\caption[a]{The $(\alpha,\nu)$ projection of the parameter 
space of the warped solution is illustrated. The curves correspond 
to lines of constant $\mu$. }
\label{F10}
\end{figure}

\begin{figure}
\centerline{\epsfxsize = 11 cm  \epsffile{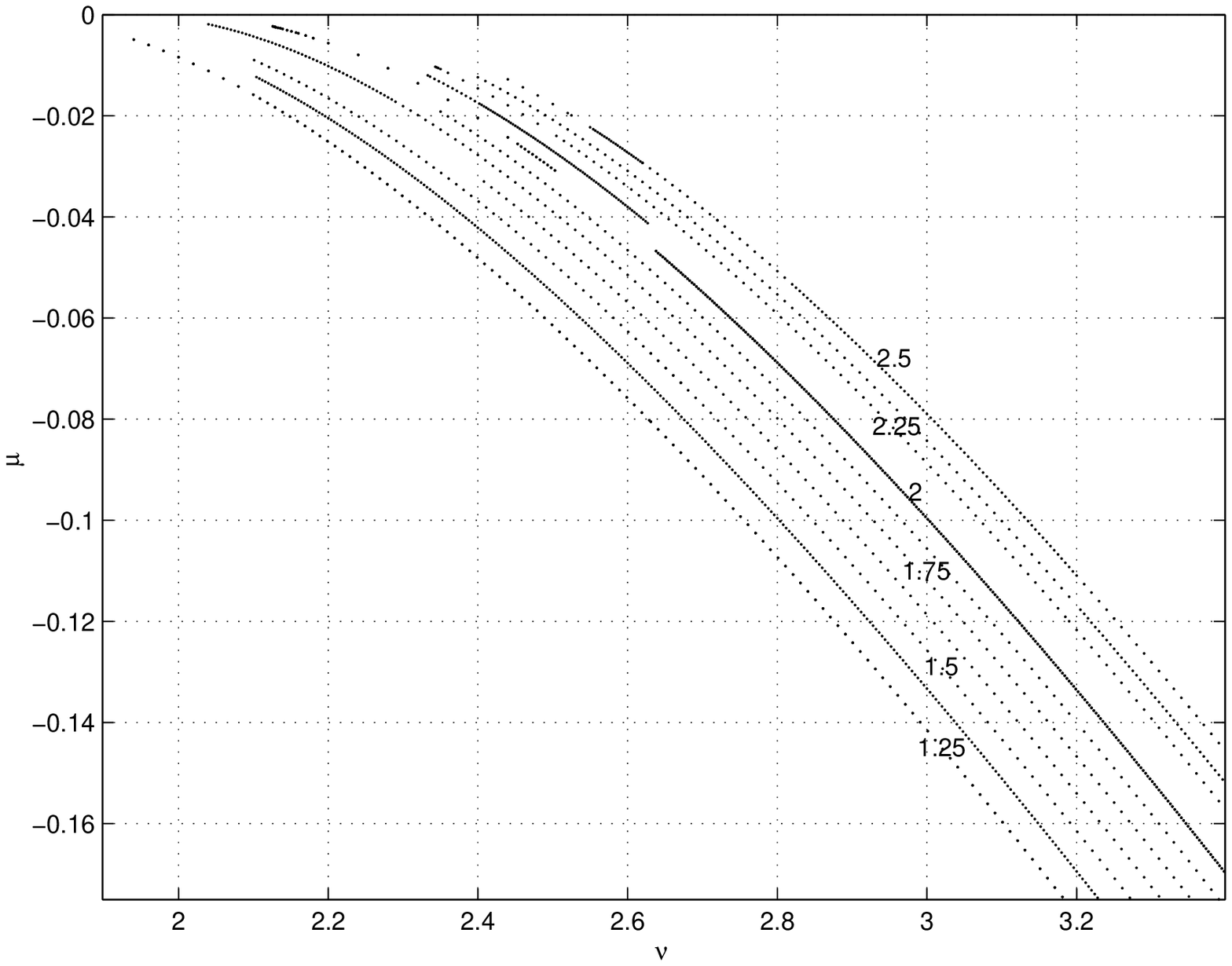}} 
\caption[a]{We report the variation of $(\mu,\nu)$ for fixed 
$\alpha$. As in the previous figures the numerical values represent 
the values of $\alpha$ which is kept constant along each curve.}
\label{F11}
\end{figure}

The same type of procedure can be carried on in the case of fixed 
$\mu$. 
As a result we will get the $(\alpha, \nu)$ projection 
of the parameter space of the solution. This result is
presented in Fig. \ref{F10}. 
Finally we can also investigate the $(\mu,\nu)$  plane at 
for fixed $\alpha$. In this case the results of our study 
are shown in Fig. \ref{F11}.
In Fig. \ref{F11} we can notice that the limit $\mu\rightarrow 0$  at
$\alpha=2$ implies that $\nu \rightarrow 2$. This can be shown
analytically,  see  Appendix.

\begin{figure}
\centerline{\epsfxsize = 11 cm  \epsffile{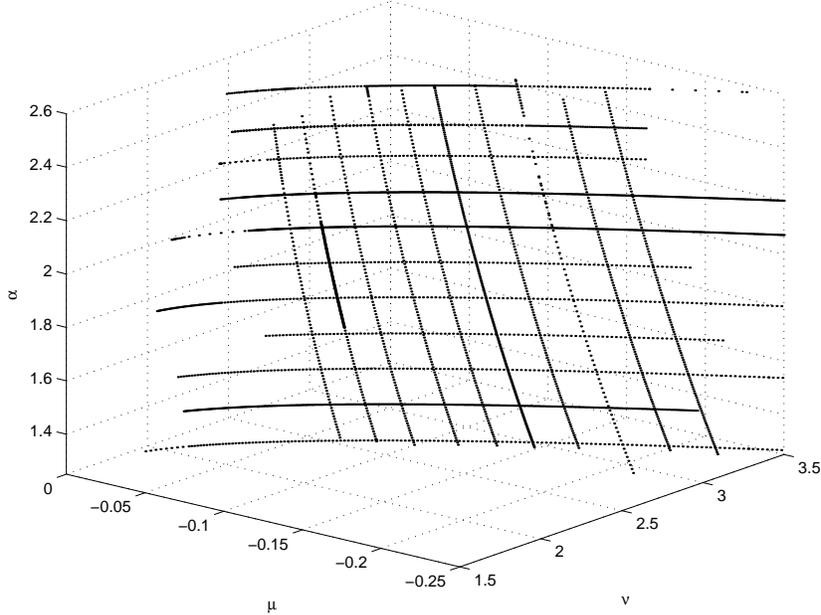}} 
\caption[a]{We illustrate the three-dimensional picture of the 
fine-tuning surface in the $(\alpha,\mu,\nu)$ plane. The points on the 
surface correspond to warped solutions obeying the boundaries
of Eq. (\ref{boundary}). }
\label{F12}
\end{figure}

Finally, in order to give a visual feeling of the fine-tuning 
surface we can present a three-dimensional plot in the 
$(\alpha,\mu,\nu)$ plane. This surface is illustrated 
in Fig. \ref{F12}.
The points on this surface 
defines a fine-tuned solution of the types 
presented in Fig. \ref{F4} and Fig. \ref{F7}.

\subsection{The thin-string limit}
The thin string limit was used for gravity localization in \cite{gs}.
In this limit  the energy density of the vortex gets more and more
localized  in the origin while the exponential damping of the  warp
factors gets milder as in Fig. \ref{F7}. Here we will show how the
thin-string limit implemented in the Abelian Higgs model.

The main idea of a thin string limit is to replace the real thick string
made of the scalar and gauge fields by a thin 
gravitating object \footnote{ The physical size
of this gravitating object,
$l_0$, shall be  much smaller than the ``size" of extra dimensions given by
$r_0= \frac{\sqrt{2}}{c m_H}=\sqrt{\frac{10 M_6^4}{8\pi\Lambda}}$.}
 characterized by three different string tensions
defined in Eq. (\ref{tension}). Then the parameters of the metric
outside the string $M_0,~{\cal L}_0,~\epsilon$ and four-dimensional
Planck constant are related in a simple way to the string tensions
[i.e. $\mu_0, \mu_\rho$ and $\mu_\theta$], to the 
 six-dimensional Planck constant,
and to the bulk cosmological constant.  These relationships were worked
out for the gravity localizing solution in Ref.\cite{gs} (see also
\cite{PP}), and they can be reproduced as well from Eqs.
(\ref{rela})--(\ref{relb}) assuming that $ c \ll 1$, which is a
condition for a thin string limit in dimensionless variables. 

Here we will write somewhat more general equations valid also for 
$\epsilon \geq 0$ or $\epsilon < -1$ that lead to regular geometry at
infinity. For $c \ll 1$ the integrals and limits $x_c \to \infty$ can
be explicitly taken, and one gets 
\begin{eqnarray}
&& M_0^4{\cal L}_0 c(1-\epsilon)^2 + {\cal O}(c^2)=
-\frac{\nu}{4}\left(\mu_\rho+\mu_{\theta}\right)~,
\label{relan}\\ 
&& M_0^4{\cal L}_0 c(1+8\epsilon +\epsilon^2)+ {\cal O}(c^2)=1-\nu
\left(\mu_0-\frac{3}{4}\mu_{\theta}
+\frac{\mu_\rho}{4}\right)~.
\label{relbn} 
\end{eqnarray}

\begin{figure}
\centerline{\epsfxsize = 11 cm  \epsffile{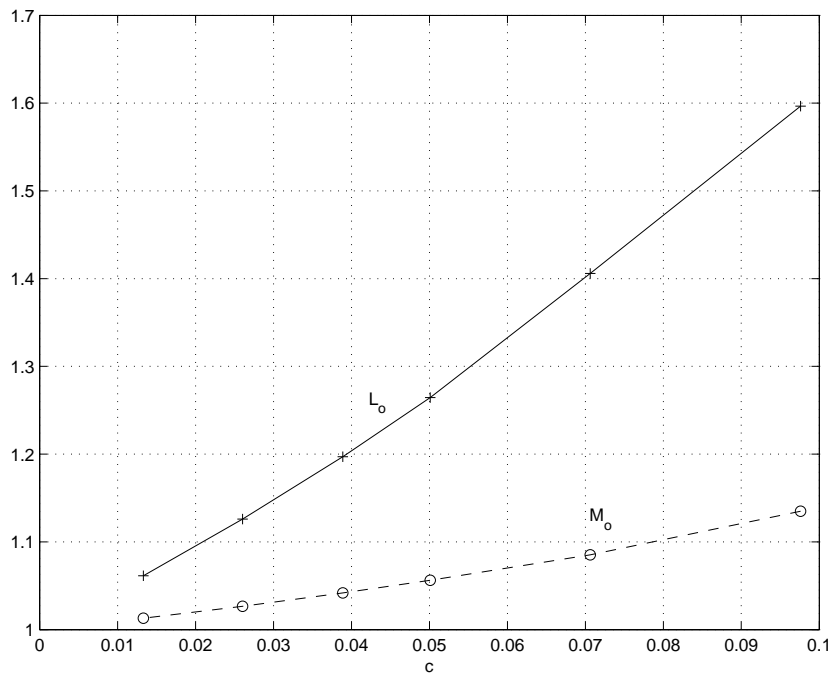}} 
\caption[a]{The variation of $M_{0}$ and ${\cal L}_0$  as a function
of $c$ for $\alpha=2$.}
\label{F19}
\end{figure}

As the numerical analysis shows (see also analytic discussion in
Appendix), in the limit $c \to 0$ the parameters $M_0$ and ${\cal
L}_0$ go to some non-zero constants (see, e.g. Fig. \ref{F19}). Then
it is clear from Eqs.  (\ref{relan})--(\ref{relbn}), that the  
combination $\mu_{\rho} + \mu_{\theta}$ scales as ${\cal O}(c)$.  In
fact, as shown in \cite{PP}, this feature has a more general
character  and it is not related to the specific form of the 
Lagrangian of the Abelian-Higgs model. 

It is interesting to ask the question whether a thin string limit
$c\ll 1$ can be organised in a way that :\\
(i) the four-dimensional Planck scale is a finite constant;\\
(ii) corrections to the four-dimensional Newton law stay small;\\
(iii) classical gravity is applicable in the bulk;\\
and, eventually \\
(iv) classical gravity is applicable in the core of the string.

In order to answer these questions let us go back to the dimension-full
parameters of the model. They are 
\begin{equation}
\Lambda\equiv M_\Lambda^6,\,\,\,M_{6},\,\,\, m_{H},\,\,\, \lambda,
\,\,\,e^2. 
\end{equation}
The other  parameters can be expressed as a function of the previous
ones  since we know that, for instance,  
\begin{equation}
\alpha = \frac{e^2}{\lambda} = 2 \frac{m_{V}^2}{m_{H}^2}~.
\end{equation}
In order to have only two scales in the problem,
we can then take $\alpha$ to be fixed to given value 
(for instance $\alpha=2$). Once $\alpha$ has been fixed, we can see if, 
in the limit $c
\ll 1$, the requirements (i)--(iv) can be satisfied. In order 
to solve the problem,
we should know the values of $M_0$ and $L_0$  defined in Eq.
(\ref{eps=0}) in the thin string limit and also the limit of
parameter $\nu$. An analytic discussion of these questions is
contained in Appendix. Here we simply state the results. In the 
limit $c\rightarrow 0$ and for $\alpha =2$, $\nu=2$ [see Fig.
\ref{F11}], and $M_0 = {\cal L}_0 = 1$ [see Fig. \ref{F19}].

(i) The four-dimensional Planck mass
\begin{equation}
M_{P}^2 = \frac{4 \pi M^4_{6}}{m_{H}^2} \int dx M^2(x) {\cal L}(x) 
\end{equation}
for small $c$ and $\alpha=2$ is simply 
\begin{equation}
M_{P}^2 = \frac{\sqrt{10\pi}}{3}\frac{M^{6}_{6} }{m_{H}
\sqrt{|\Lambda|}}~,
\end{equation}
which must be equal to $[1.22 \times 10^{19} {\rm GeV}]^2$.

(ii) Corrections to 4d gravity were computed in \cite{gs} and are
given by ${4 \over 3\pi}(r_0/r)^3$. They are known to be small at the distances
smaller than $0.2$ mm \cite{submm} and thus we require
\begin{equation}
r_0 = \frac{0.2~{\rm mm}}{\xi}~,~~ \xi > 1~.
\end{equation}

(iii) To parameterize the Planck constant in the bulk, we write
\begin{equation}
M_6= \eta \, 10^3 {\rm GeV}~. 
\end{equation}
The choice of $\eta \sim 1$ makes the fundamental Planck scale to be
of the order of the weak scale and may potentially lead to the
solution of the gauge hierarchy problem \cite{dim}. To satisfy the third
requirement, we must have the curvature in the bulk to be smaller than
$M_6^2$, i.e. $M_\Lambda < M_6$.

(iv) If we require the validity  of gravity  on the string we should
demand that the (quadratic) curvature  invariants of the  solution are
always much smaller than $M_{6}^4$. In the small $x$ limit the curvature
invariants are 
\begin{eqnarray}
&& R^2 \sim m^4_{H} \, \frac{\,{\left( -8\,\alpha + 
         \nu\,\left( 4\,\mu + \nu \right)  \right) }^2}
     {4\,{\nu}^2},
\nonumber\\
&& R_{AB}R^{AB} \sim m^4_{H}\, \frac{5\,{\left( -8\,\alpha + 
         \nu\,\left( 4\,\mu + \nu \right)  \right) }^2}
     {64\,{\nu}^2},
\nonumber\\
&& R_{ABCD}R^{ABCD} \sim m^4_{H}\,
\frac{{\left( -8\,\alpha + 
        \nu\,\left( 4\,\mu + \nu \right)  \right) }^2}
     {16\,{\nu}^2}, 
\nonumber\\
&&C_{ABCD} C^{ABCD} \sim m_{H}^4\,
\frac{3\,{\left( -8\,\alpha + 
         \nu\,\left( 4\,\mu + \nu \right)  \right) }^2}
     {320\,{\nu}^2}. 
\end{eqnarray}
So,  we find that this requirement is satisfied if
\begin{equation}
\frac{m_H}{M_6} \ll 1~. 
\end{equation}

In numbers, the thin string limit $c \ll 1$ implies
\begin{equation}
c \simeq 70 \times \frac{\xi^2}{\eta^4} \ll 1~,
\end{equation}
the third condition requires
\begin{equation}
\frac{\eta}{\xi} \gg 6 \times10^{-16},
\end{equation}
and the ratio $\frac{m_H}{M_6}$ essential for the requirement (iv) is
\begin{equation}
\frac{m_H}{M_6} \simeq 2 \times 10^{-17} \frac{\eta^3}{\xi}~.
\end{equation}
One can see that all these conditions can be satisfied in a wide
range of the parameters of the model. For example, the hierarchy $M_6
\ll M_{P}$ can be easily achieved for $\eta \sim 10,~~\xi \sim 1$.

If the limit $c \to 0$ is taken in mathematical sense, it is
impossible to satisfy all four conditions, because the curvature of
the space inside the string diverges. However, the first three
conditions can be satisfied, if the parameters of the model scale as:
\begin{eqnarray}
&&m_H =E \biggl({1\over c}\biggr)^{\beta}~,
\nonumber\\
&&\Lambda = E^6 \biggl({1\over c}\biggr)^{4\beta-3},
\nonumber\\
&& M_{6}^4 = E^4  \biggl({1\over c}\biggr)^{2\beta-1},
\nonumber\\
&& \lambda = E^{-2} \biggl({1\over c}\biggr),
\label{scaling}
\end{eqnarray}
with $1< \beta <{3\over 2}$ and  $E$ being an arbitrary energy scale.

\renewcommand{\theequation}{6.\arabic{equation}}
\setcounter{equation}{0}
\section{Extensions}

\subsection{Higher windings}

The parameter space of the solutions discussed in the previous sections
has been derived in the case  $n=1$. 
It is also  possible to discuss the case of higher windings 
$n \geq 2$. The study of these cases 
can be performed using the same techniques discussed in the case of 
$n=1$. There are obvious differences 
in the asymptotics of the solutions. 
For the gauge field we have to require that $P(0) = n$. This 
imposes a different asymptotic solution 
in the core, namely
\begin{eqnarray}
f(x)&=&Ax^n+Fx^{n+2}\label{hw1},\\
P(x)&=&n+Bx^2,\\
M(x)&=&1+Cx^2,\\
{\cal L}(x)&=&x+Dx^3.
\label{hw2}
\end{eqnarray}
For $n\geq 2$, we find:
\begin{eqnarray}
C&=&-\frac{\mu}{8}-\frac{\nu}{32}+\frac{B^2\nu}{4\alpha},\\
D&=&\frac{\mu}{12}+\frac{\nu}{48}-\frac{5B^2}{6\alpha},\\
F&=&A\left(-1+2Bn-\frac{B^2\nu}{3\alpha}(n-5)+\frac{5n-1}{6}
\left(\mu+\frac{\nu}
{4}\right)\right).
\label{hw2b}
\end{eqnarray}
With these solutions it is possible to integrate the 
system by repeating 
the same procedure discussed in the case $n=1$. 

Since the winding number determines the value of the 
gauge field at the origin the relations among the 
string tensions will also be modified. In particular, Eq. 
(\ref{in3}) becomes, in the case of $n \geq 2$, 
\begin{equation}
\mu_0-\mu_\theta=
-\frac{n}{\alpha}\left.\frac{P'}{\cal L}\right|_0,
\label{hw3}
\end{equation}
which implies that $B$, appearing in Eqs. (\ref{hw1})--(\ref{hw2}),
is now 
\begin{equation}
B = -\frac{\alpha}{2 n \nu}~.
\label{hw4}
\end{equation}
As in the case $n=1$ the tuning of the string tensions 
expressed in terms of $B$ is only necessary (but not sufficient) 
in order to get 
solutions which lead to gravity localization. Thus, as in the case 
$n=1$, $A$ should be again tuned. 
\begin{figure}
\centerline{\epsfxsize = 11 cm  \epsffile{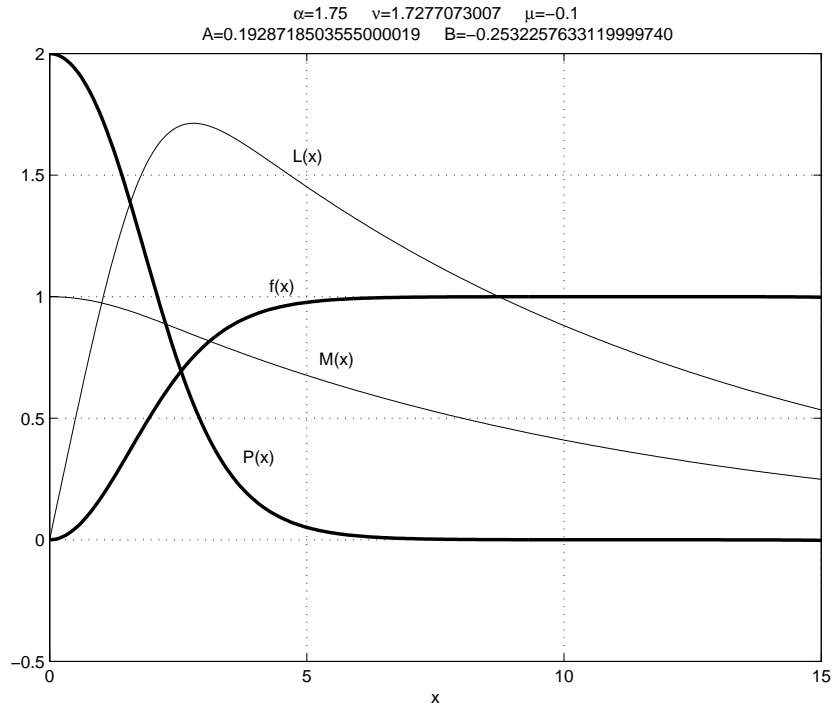}} 
\caption[a]{A solution leading to gravity localization
in the case of winding $n=2$ is illustrated.}
\label{F20}
\end{figure}
In Fig. \ref{F20} we illustrate an example of regular solutions 
with $n = 2$. The components of the energy-momentum 
tensor corresponding to the solution of Fig. \ref{F20} are 
reported in Fig. \ref{F20b}.
\begin{figure}
\centerline{\epsfxsize = 11 cm  \epsffile{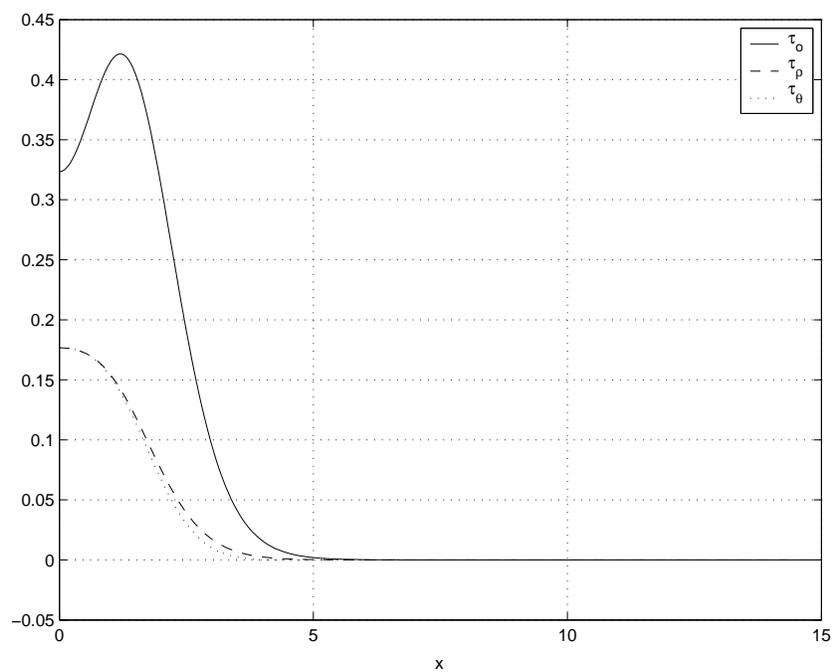}} 
\caption[a]{The components of the energy-momentum tensor 
computed from the solution of Fig. \ref{F20} are illustrated.}
\label{F20b}
\end{figure}

\begin{figure}
\centerline{\epsfxsize = 11 cm  \epsffile{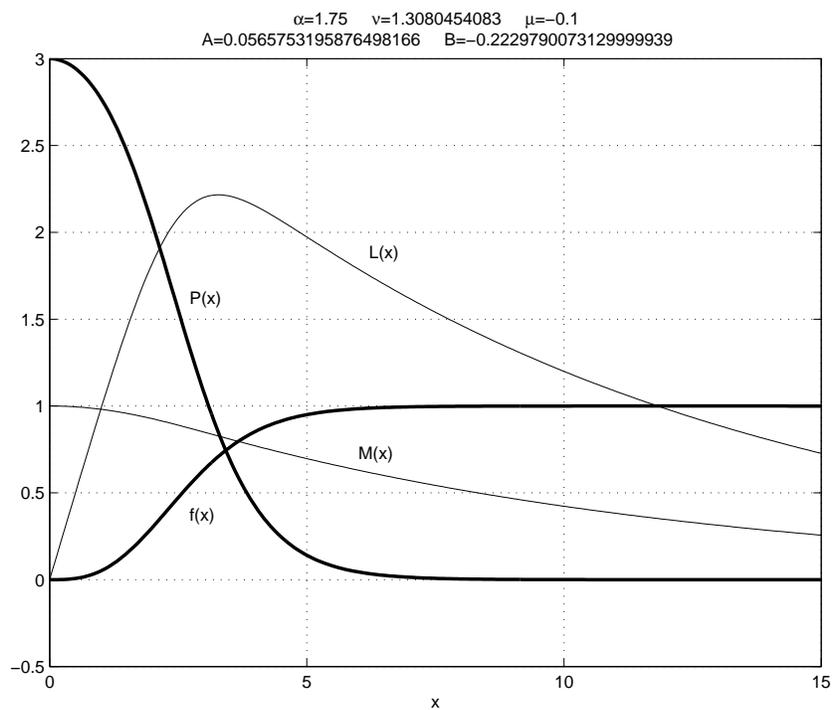}} 
\caption[a]{An example of solution with $n =3$. }
\label{F21}
\end{figure}

\begin{figure}
\centerline{\epsfxsize = 11 cm  \epsffile{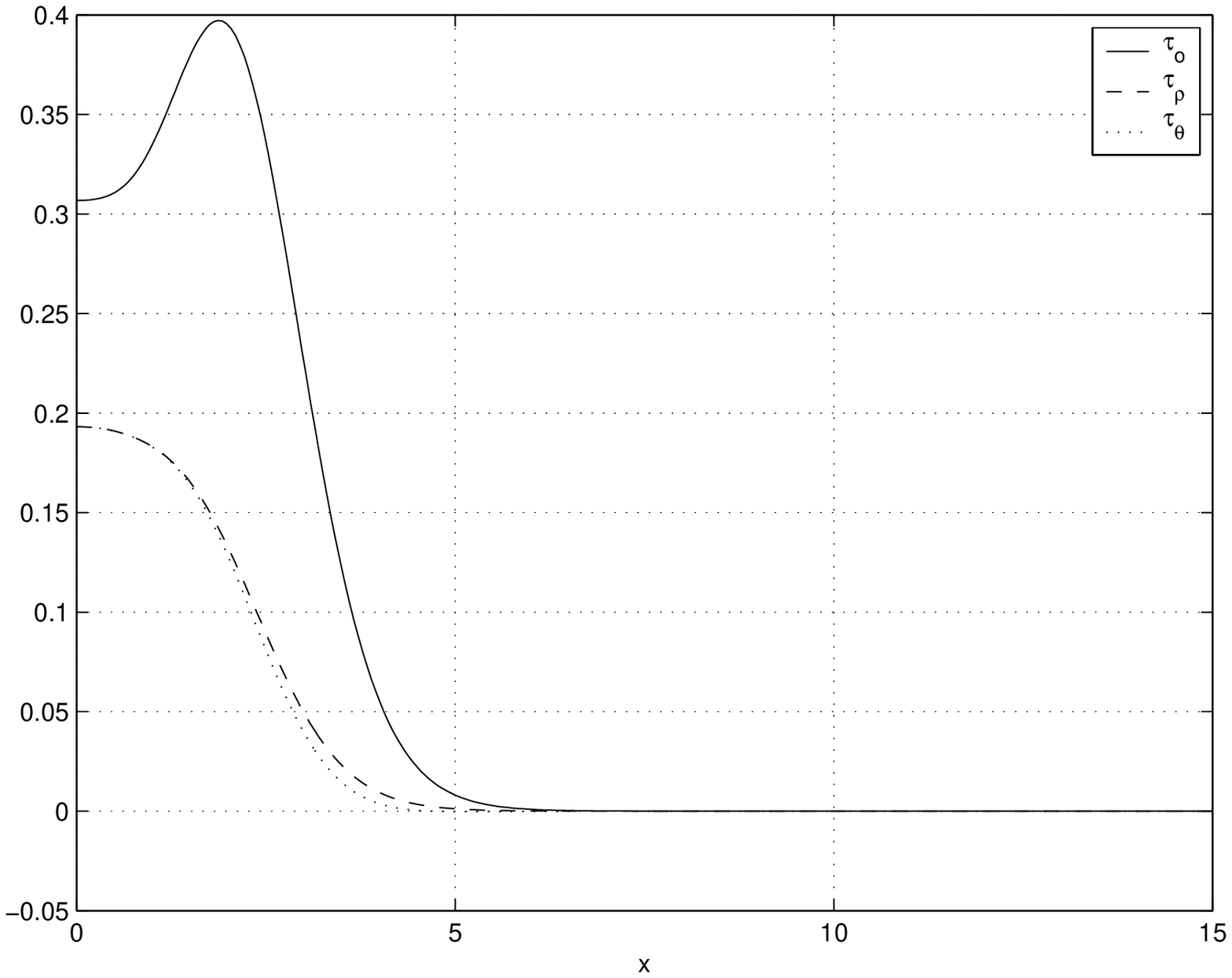}} 
\caption[a]{The components of the energy-momentum tensor in the case 
of the solution reported in Fig. \ref{F21}. }
\label{F21b}
\end{figure}

In Fig. \ref{F21} a solution which localizes gravity is reported in the case 
$n =3$.  
In Fig. \ref{F21b} the components of the energy-momentum tensor are 
reported for the solution illustrated in Fig. \ref{F21}.
The solutions illustrated in Fig. \ref{F20} and \ref{F21} lead to regular 
geometry since all the curvature invariants are finite for any $x$.
As in the case $n =1$ a full scan of the parameter 
space could be performed following the same procedure 
outlined in the body of the paper. It is interesting 
to speculate that these solutions may have some implications 
in models trying to explain the  origin of families 
in a higher dimensional context \cite{tro}. It would then be of some 
interest to study, in the context of solutions with higher 
windings, the structure and localization of the fermionic 
zero modes.

\subsection{String solutions without localization of gravity}

The examples discussed up to this point illustrated mainly 
 solutions whose warp factors  decrease at infinity. In this section we will
give some examples of solutions with the opposite type of behaviour.

\begin{figure}
\centerline{\epsfxsize = 11 cm  \epsffile{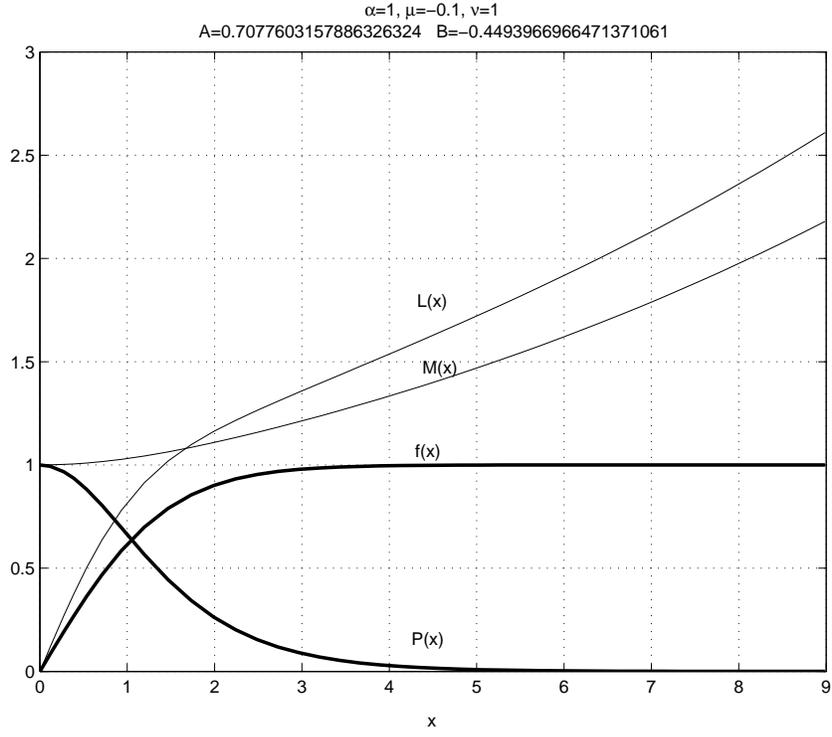}} 
\caption[a]{We illustrate and anti-warped solution with parameters 
$\alpha =1$, $\nu =1$ and $\mu = -0.1$.}
\label{F13}
\end{figure}

\begin{figure}
\centerline{\epsfxsize = 11 cm  \epsffile{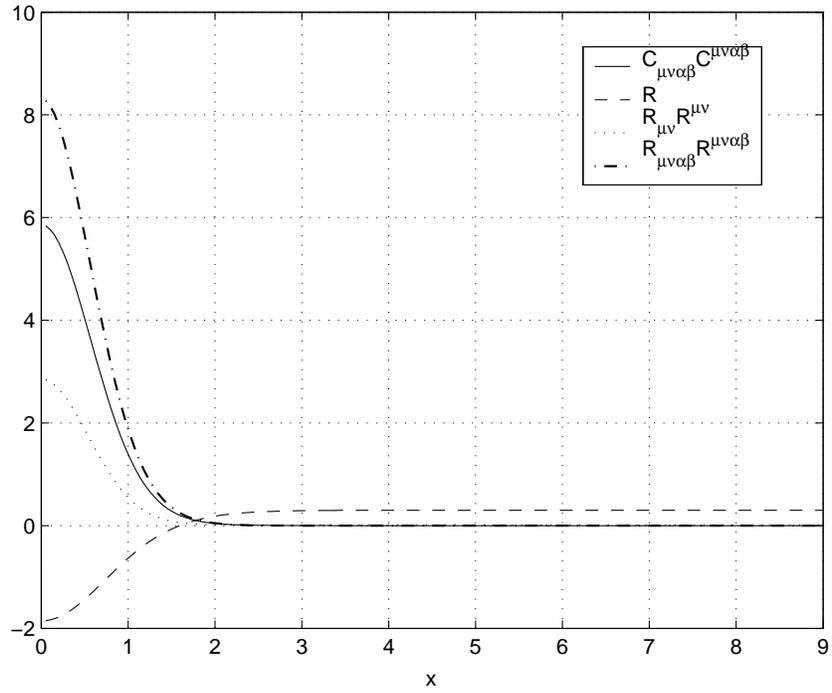}} 
\caption[a]{The  curvature invariants for the solution shown in Fig.
\ref{F13}. } 
\label{F13b}
\end{figure}

In Fig. \ref{F13} we show a regular solution of Eqs.
(\ref{f1})--(\ref{l1})  where the warp factors increase at large
distance from the core. The parameters of Fig. \ref{F13} are $\alpha
=1$, $\nu =1 $,  $\mu = -0.1$.
In the solution of Fig. \ref{F13} Eq. (\ref{B}) does not hold. From Eq.
(\ref{B}) we should have $B = -0.5$ whereas, for Fig. \ref{F13} 
$B\neq -0.5$. In this sense the solution is not fine-tuned. In 
Fig. \ref{F13b} we plot the curvature invariants pertaining 
to the solution of Fig. \ref{F13}.
For large $x$ the asymptotic behaviour illustrated  in Fig. \ref{F13}
can be related to the solutions given in Eq. (\ref{mm}). 
 
\begin{figure}
\centerline{\epsfxsize = 11 cm  \epsffile{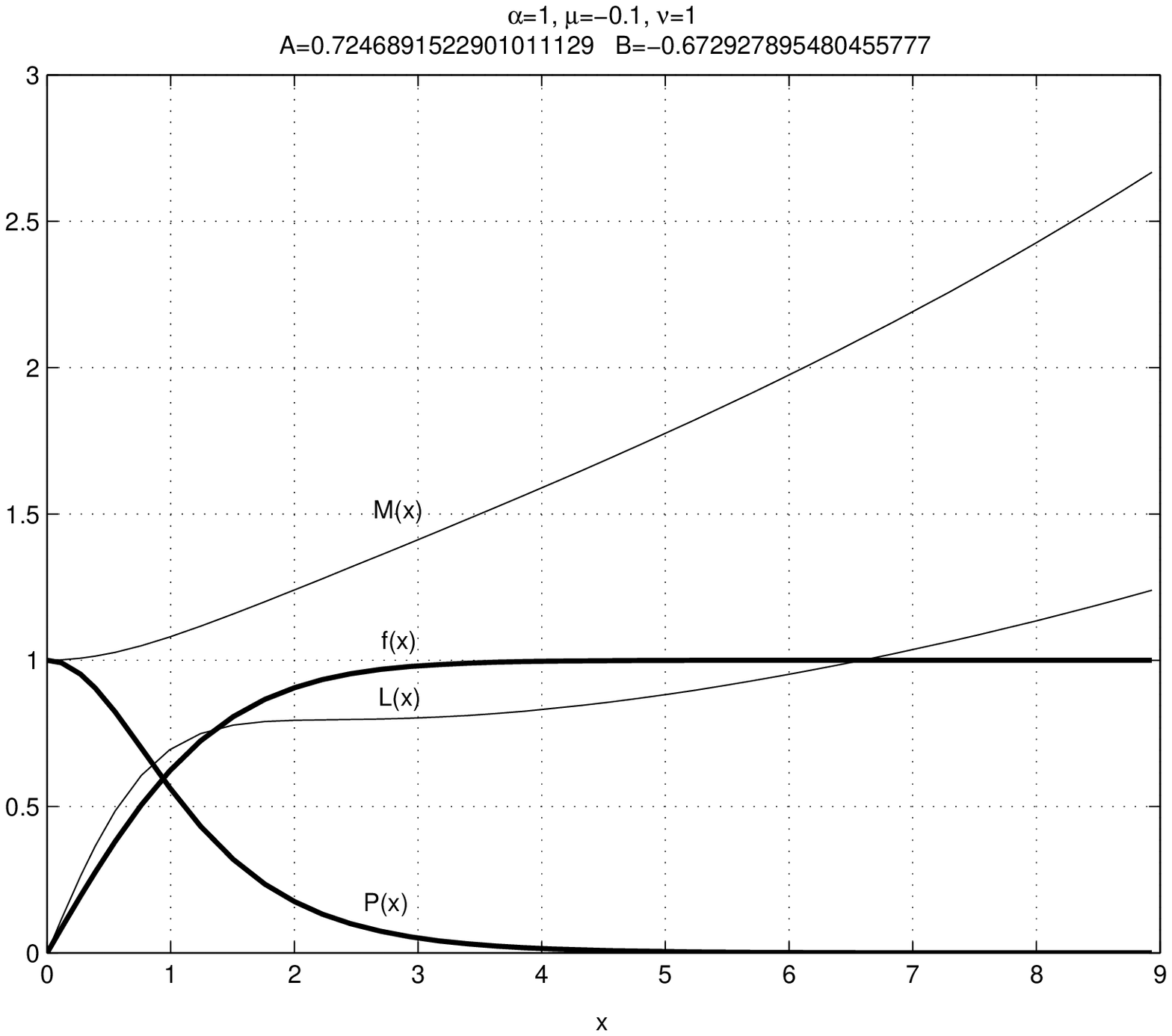}} 
\caption[a]{The parameters of the solution discussed 
in this plot are the same of Fig. \ref{F13} but with different $A$ and $B$. 
The curvature 
in the origin is larger than in the case of Fig. \ref{F13b}. }
\label{F14}
\end{figure}

\begin{figure}
\centerline{\epsfxsize = 11 cm  \epsffile{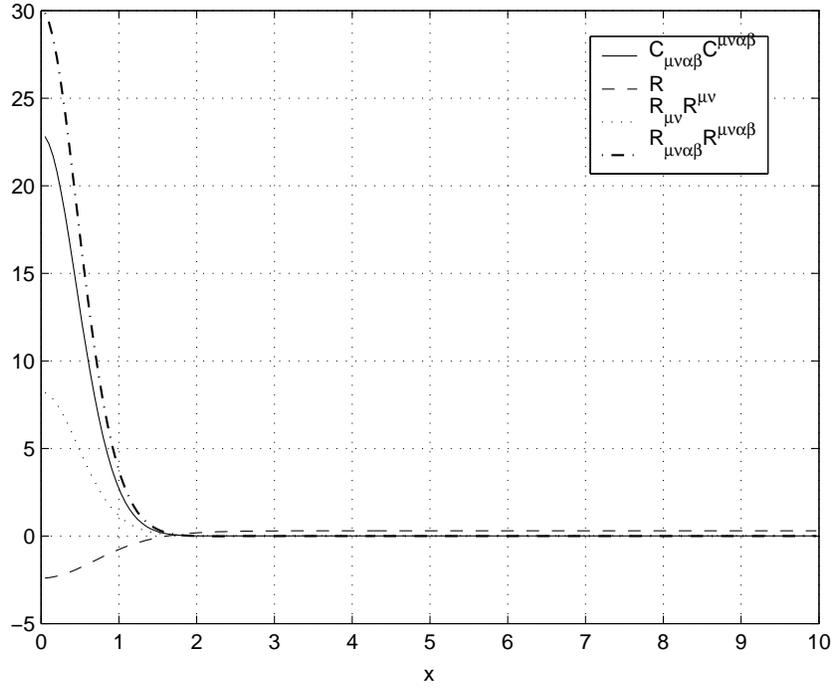}} 
\caption[a]{The curvature invariants for the solution 
of Fig. \ref{F14}. }
\label{F14b}
\end{figure}

It is interesting that the same values of $\alpha$, $\nu$ and  $\mu$
selected for the solution of Fig. \ref{F13} admit a physically
distinct extra solution. It is  shown in Fig. \ref{F14} and the
related curvature  invariants  are illustrated in Fig. \ref{F14b}.
We will not attempt here a full classification of the solutions. We can notice,
however, that the solutions illustrated in Fig. \ref{F13} 
and Fig. \ref{F14} can be distinguished (in a coordinate independent 
fashion) from the behaviour of the curvature invariants in the 
origin. As we can see from Figs. \ref{F13b} and \ref{F14b} 
the solution of Fig. \ref{F14} leads to larger curvature invariants 
in the origin.

\begin{figure}
\centerline{\epsfxsize = 11 cm  \epsffile{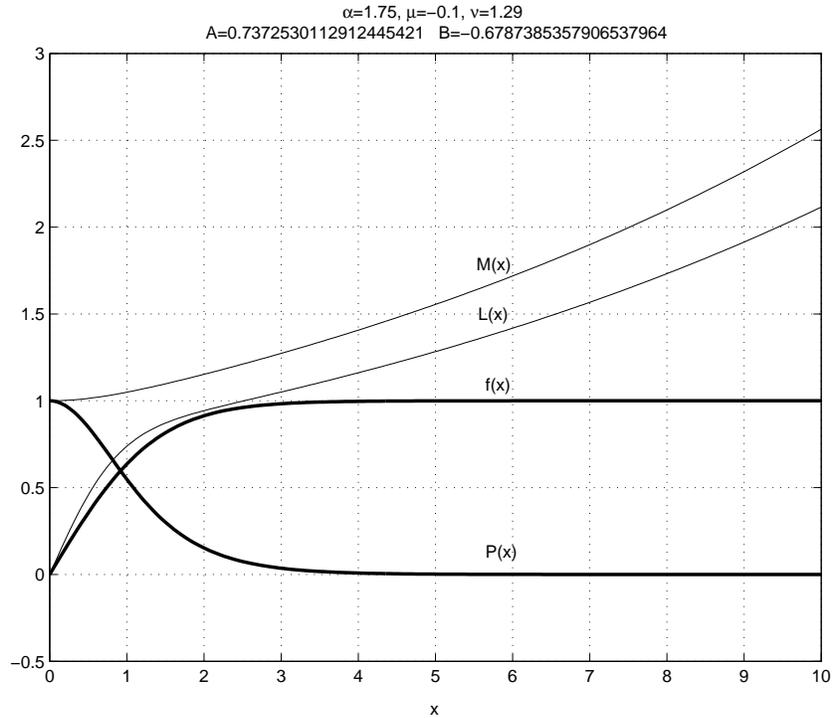}} 
\caption[a]{We illustrate a solution where the 
string tensions have been 
fine-tuned according to Eq. (\ref{B}) but with the metric growing at
infinity.}
\label{F15}
\end{figure}

\begin{figure}
\centerline{\epsfxsize = 11 cm  \epsffile{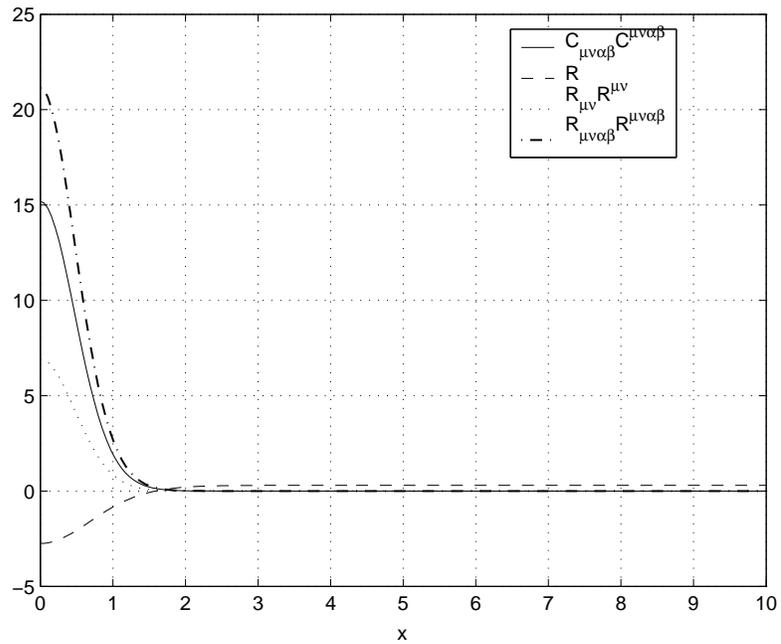}} 
\caption[a]{The curvature invariants associated with 
the solution of Fig. \ref{F15}. }
\label{F15b}
\end{figure}

Finally we want to demonstrate that the tuning of the string
tensions  as implied by Eq. (\ref{B}) is not a sufficient condition 
in order to obtain solutions with localization of gravity. In Fig.
\ref{F15} and Fig. \ref{F15b} we illustrated  a regular solution whose parameters are
given by $\alpha=1.75$, $\nu =1.29$ and $\mu = -0.1$ and tensions
fine tuned according to Eq. (\ref{B}). Nevertheless, the metric
increases at infinity.
Asymptotically for large $x$ the solution of Fig. \ref{F15} 
are the ones of Eq. (\ref{mm}) with $\epsilon \to \infty$.

\section{Concluding remarks}

In this paper we discussed the possibility  of obtaining 
solutions which localize gravity in six-dimensions.  The source of
the gravitational field has been  assumed  to be a generalized
version of the Nielsen-Olesen vortex  living in the two-dimensional
transverse space.

We show that the localization of gravity is possible on a ``thick"
string and found a fine-tuning condition leading to a set of
physically interesting solutions. Since the geometries described in
this paper are regular (i.e. curvature singularities  are absent)
gravity can be described in classical terms both in the  bulk and on
the vortex. We studied a thin string limit and identified the choice
of parameters that may potentially lead to a solution to the gauge
hierarchy problem. 

Various questions are still open. Our explicit solutions  couple
together scalar, tensor and gauge degrees of freedom. Therefore  they
represent an ideal  framework where  the localization of fields of
various spin can be explicitly analyzed in  a completely regular
geometry. There are also open questions concerning  higher windings.
It has been shown that also for higher windings  there are solutions
which localize gravity. It would be natural  to ask if these
solutions are as stable as the ones obtained in the  case of the
lowest winding.  Finally, since the geometries discussed  in this
paper are completely regular (in a technical sense) it is
interesting  to understand their stability against first order
(quadratic) corrections to the Einstein-Hilbert action \cite{max}
(which may naturally arise in a string theoretical context).

We thank P. Tinyakov and K. Zuleta for discussions. This work was
supported by the FNRS grants no.  21-55560.98, 7SUPJ62239
and by the Tomalla Foundation.

\newpage
\begin{appendix}

\renewcommand{\theequation}{A.\arabic{equation}}
\setcounter{equation}{0}
\section{The Bogomolnyi limit}

\begin{figure}
\centerline{\epsfxsize = 11 cm  \epsffile{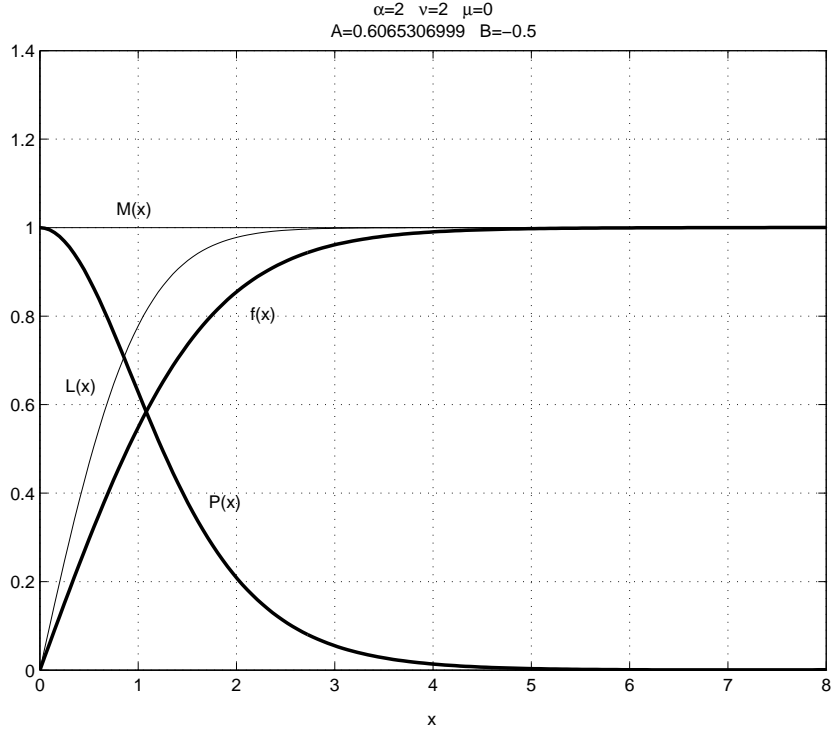}} 
\caption[a]{A solution in the Bogomolnyi limit for $c \ll 1$. }
\label{F8}
\end{figure}

In this appendix we show analytically that for the case $\alpha=2$
(the Bogomolnyi limit \cite{bog}) the following limits hold:
\begin{eqnarray}
&&\lim_{c\rightarrow 0} \nu(c) = 2~,
\label{NU}\\
&&\lim_{c\rightarrow 0} M_0(c) = 1~,
\label{limit2}
\end{eqnarray}
and 
\begin{equation}
\lim_{c\rightarrow 0} {\cal L}_0(c) =1~,
\label{llim}
\end{equation}
if the string tensions are tuned according to Eqs. 
(\ref{in2})--(\ref{in3}). The numerical solution for the Bogomolnyi
limit is shown in Fig. \ref{F8}.

In the Bogomolnyi limit and for $c=0$ the equations of motion for the
gauge and scalar fields reduce to a coupled set of first 
order non-linear differential equations which can be written as 
\begin{eqnarray}
&&{\cal L}\,f'=f\,P,
\nonumber\\
&&\frac{P'}{\cal L}= f^2-1.
\label{bog}
\end{eqnarray}
If we insert Eqs. (\ref{bog}) into Eqs. (\ref{t0})-(\ref{tth}) we obtain 
that 
\begin{equation}
\tau_{\rho}(x)= \tau_{\theta}(x) =0.
\end{equation}
The equation for  $M$ becomes then
\begin{equation}
4 m' + 10 m^2 = - \mu.
\label{bog2b}
\end{equation} 
The limit $c\rightarrow 0$  implies  that $\mu \rightarrow 0$. 
From Eq. (\ref{bog2b}) the 
only (regular) solution respecting the boundary conditions 
at the origin is then $M(x)=M_0= 1$. Thus, this proves Eq.
(\ref{limit2}).

Since $\tau_{\rho}(x)=0$, from Eq. (\ref{tx=0}) we get $B^2 =
\alpha/8=1/4$ and from Eq. (\ref{B}) $\nu =2$. Thus, this proves
that, once the string tensions are tuned, Eq. (\ref{NU}) is valid.

Now, the difference of Eqs. (\ref{m1}) and (\ref{m2}) yields,
\begin{equation}
(M^3M'L)'-(M^4L')'=\nu M^4L(f_o-f_\theta)~.
\label{!}
\end{equation}
By using  for $P(x)$ Eq. (\ref{p1}) we have: 
\begin{equation}
M^4L(f_o-f_\theta)=\left(\frac{PP'M^4}{\alpha L}\right)'~.
\end{equation}
Thus all terms in (\ref{!}) are total derivatives; integrating 
both sides from 0 to $x$ we obtain
\begin{equation}
m-\ell=\frac{\nu}{\alpha}\frac{PP'}{L^2}-\left(1+\frac{\nu}{\alpha}n\left.
\frac{P'}{L}\right|_0
\right)\frac{M_o^4}{M^4L}~,
\label{last}
\end{equation}
where $M_o$ is the value of $M$ on the core of the string and $n$ is 
the winding. In the case of fine-tuned solution, according to Eq.
(\ref{B}), the last term in Eq. (\ref{last}) vanishes.
Since we just showed that in the limit $\alpha \to 2$ we have 
$\nu \to 2$ and $M \to 1$, from Eq. (\ref{last}) we get
\begin{equation}
L L' = - PP'~,
\label{la}
\end{equation}
which gives, after integration over $x$,
\begin{equation}
L^2=1-P^2~,
\end{equation}
where we used the fact that $P(0)=1$.
This proves (\ref{llim}), because $P(x) \to 0$ for $x \to \infty$.

\end{appendix}

\end{document}